\documentclass[fleqn,10pt]{wlscirep}

\usepackage[utf8]{inputenc} 
\usepackage[T1]{fontenc} 

\usepackage{braket}

\usepackage{siunitx}
\DeclareSIUnit\gauss{G}
\DeclareSIUnit\torr{Torr}
\DeclareSIUnit\LabView{LV}
\DeclareSIUnit\pixel{px}
\DeclareSIUnit\inch{^{\prime\prime}}
\DeclareSIUnit\miliK{mK}

\newcommand{\tab}[1][1cm]{\hspace*{#1}}

\usepackage{nccmath}

\title{Developing a simulation tool to investigate a novel trapped two-state Bose-Einstein condensate Ramsey interferometer driven by dipole oscillations and gravitational sag}

\author[1,2,*,+]{Anushka Thenuwara}
\author[1,+]{Andrei Sidorov}
\affil[1]{Centre for Optical Sciences, Faculty of Science, Engineering and Technology, Swinburne University of Technology, John Street, Hawthorn, Victoria, Australia 3122}
\affil[2]{School of Physics and Astronomy, Monash University Clayton Campus, Wellington Rd, Clayton, Victoria, Australia 3800
}

\affil[*]{anushka.thenuwara@monash.edu}

\affil[+]{these authors contributed equally to this work}

\begin{abstract}
We propose and explore the feasibility of a novel Ramsey interferometer created by a trapped two-state Bose-Einstein condensate (BEC) driven by dipole oscillations and gravitational sag. The BEC is formed in a pure cigar shaped compressed magnetic trap (CMT) via a dilute atom cloud of $^{87}Rb$ atoms in state $\ket{F=2, m_F=+2}$ $(\ket{+2})$ of the $5 ^2S_{\frac{1}{2}}$ ground state. Here, Rmasey interferometry is performed with states $\ket{F=2, m_F=+1}$ $(\ket{+1})$ and $\ket{+2}$. The proposed interferometer utilises the response of atoms to the harmonic oscillator trapping potential and the gravitational sag due to the variation in the $m_F$ state. Briefly, the state $\ket{+1}$ experiences a shallower radial trap with a larger gravitational sag; whereas, state $\ket{+2}$ experiences a tighter radial trap with a gravitational sag which is half of state $\ket{+1}$. Due to this, a superposition between the states $\ket{+1}$ and $\ket{+2}$ experiences multipath propagation resulting in an interference pattern. This may be utilised to measure local gravitational fields and measure inter-sate scattering lengths. Here, a theoretical framework is reported which is developed via the two-level system in combination with the Gross-Pitaevskii equation (GPE). Further, the development of a simulation tool via GPELabs in MATLAB that explores the prosed interferometer is reported along with key insights and findings.
\end{abstract}
\begin{document}

\flushbottom
\maketitle
%
%
\thispagestyle{empty}

\section*{Introduction}
The $5 ^2S_{\frac{1}{2}}$ ground state of $^{87}Rb$ atoms have three magnetically trappable states: $\ket{m_F = +2} (\ket{+2})$ and $\ket{m_F = +1} (\ket{+1})$ in the $F = 2$ hyperfine manifold and $\ket{m_F = -1} (\ket{-1})$ in the $F = 1$ hyperfine manifold. Two of those states ($\ket{+1}$ and $\ket{-1}$) have the same magnetic moment, can be coupled by a two-photon MW-RF transition and have been extensively studied \cite{Deutsch2010, Bernon2013, Srkny2014}. Moreover, several examples of Ramsey interferometry in the $\ket{-1} - \ket{+1}$ system are presented in \cite {Anderson2009, Egorov2011, Opanchuk2012, Egorov2013}. The work in \cite{Anderson2009} shows the spatially dependent relative phase evolution of an elongated two-component Bose-Einstein condensate between the states $\ket{-1}$ and $\ket{+1}$. In it, the relative phase evolution is probed via Ramsey interferometry and goes on to show that the mean-field formalism provides a good description of the decay of the Ramsey signal. It also concludes that the loss of Ramsey contrast is due to the inhomogeneity of the collective relative phase across the cloud rather than to decoherence or phase diffusion. Ramsey interferometry with spin-echo techniques in the $\ket{-1}-\ket{+1}$ system is performed in \cite{Egorov2011} and reports coherence times of several seconds. The cause of this extended coherence is due to mean-field-driven collective oscillations of the two components leading to periodic de-phasing and re-phasing of the condensate wavefunctions with a slow decay of the interference fringe visibility. The work in \cite{Opanchuk2012} describes the Wigner calculations developed with the experimental data in \cite{Egorov2011} that goes beyond classical GPE modelling. Finally, the work in \cite{Egorov2013} utilises Ramsey interferometry in the $\ket{-1}-\ket{+1}$ system to perform precise scattering length measurements. It reports the inter-state scattering length of $a_{\ket{-1}-\ket{+1}} = 98.006 (16)a_0$ with an uncertainty of \num{1.6e-4} and the intra-state scattering length of $a_{\ket{+1}-\ket{+1}} = 95.44 (7)a_0$. 

This paper focusses on exploring the capabilities of performing Ramsey interferometry in the $\ket{+1}-\ket{+2}$ system. The states $\ket{+2}$ and $\ket{+1}$ can be coupled by a single-photon magnetic dipole transition.They have different magnetic moments resulting in different radial trapping frequencies $\omega_{r}$ based on 
\begin{ceqn}
\begin{equation}
	\label{eqn:O_r}
	\omega_{r} = \sqrt{\frac{\mu_B m_F g_F}{m}\frac{\partial^2 B}{\partial r^2}},
\end{equation}
\end{ceqn}
where $\mu_B$ is the Bohr magneton, $m_F$ is the magnetic quantum number,  $g_F$ is the Land\'{e} factor of the $F$ hyperfine state, $m$ is the mass of the atom and $\frac{\partial^2 B}{\partial r^2}$ is the curvature of the magnetic field in the radial direction.

Further, BECs are affected by the gravity and the vertical position is shifted by the gravitational sag $z_g$
\begin{ceqn}
\begin{equation}
	\label{eqn:G_Sag}
	z_g = \frac{g}{\omega_r^2},
\end{equation}
\end{ceqn}
where $g$ is the gravitational acceleration. 

\begin{figure}[h!]
	\centering
	\includegraphics[width=0.8\linewidth]{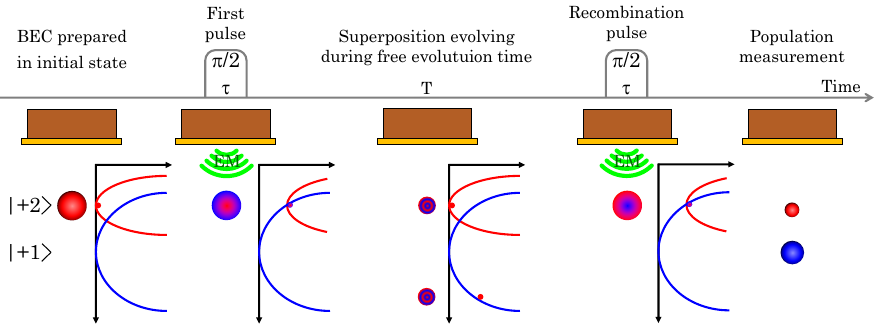}
	\caption[Ramsey sequence for $\ket{+1}-\ket{+2}$ BEC system]{Ramsey interferometry of the trapped two-state $\ket{+1} - \ket{+2}$ BEC system subjected gravitational sagging and dipole oscillations. Here, the graph next to the BEC is a schematic representation of the trapping potentials for states $\ket{+1}$ (blue) and $\ket{+2}$ (red).} 
	\label{fig:RamseySeq2}
\end{figure}

We propose a Ramsey interferometer by combining the effects of the variation in $m_F$ values on $\omega_{r}$ and $z_g$ in the $\ket{+1}-\ket{+2}$ system. Firstly, prepare a BEC in state $\ket{+2}$ and create a superposition by transfer a certain fraction of atoms to state $\ket{+1}$. This will be the first pulse of the Ramsey sequence ($\hat{U}_{\tau}$ in Equation \ref{eqn:U_pi_T}) where the experimental results in \cite{Petrovic2013} show the ability of the $\ket{+1}-\ket{+2}$ BEC system to create a variety of superpositions. Once the superpostion is created, the state dependent radial trap frequency as per Equation \ref{eqn:O_r} and state dependent gravitational sag as per Equation \ref{eqn:G_Sag} cause the $\ket{+1}-\ket{+2}$ superposition to undergo dipole oscillations. Here, the fraction of the superposition in state $\ket{+1}$ is on the slope of its trapping potential which slides down, travels towards and goes beyond its trap minimum along the $z$-axis, and oscillate back to the original position. The dipole oscillation time of $\ket{+1}$ is $T = \frac{2\pi}{\omega_{1r}}$ which is the free evolution time $T$ of the Ramsey sequence ($\hat{U}_{T}$ in Equation \ref{eqn:U_pi_T}). This free evolution time can be varied by adjusting the radial trap frequency via changing the curvature of the external magnetic field. Once the superposition undergoes one dipole oscillation, the recombination pulse $\hat{U}_\tau$ is applied with varying phase $\phi$ (Equation \ref{eqn:U_pi_T}). This completes the Ramsey interferometric sequence, where the resulting system wavefunction follows Equation \ref{eqn:Psi_t}. Based on the above, the proposed Ramsey sequence is depicted in Figure \ref{fig:RamseySeq2}.

Primarily, we explore the plausibility of creating the aforementioned  Ramsey interferometer governed by the interesting characteristics of gravitational sagging and dipole oscillations. This plausibility is theoretically explored in \cite{Thenuwara2023} where analytical expressions are derived to fit experimental results. Here, we develop and perform numerical simulations to get an insight on how the BECs behave during experimental runs. It should be noted  that these simulations follow the two-level model and only consider the trapped $\ket{+1} - \ket{+2}$ states of the  $^{87}Rb$  BEC in the hyperfine $F = 2$ manifold of the $5 ^2S_{\frac{1}{2}}$ ground state. This paper is structured in three sections, which outlines the theoretical framework, the development of the simulation tool and the results from the simulations. The section on the theoretical framework presents three areas. Firstly, the Rabi model for a two-level atom which describes the creation of a superposition of states via an electromagnetic (EM) field through the matrix formalism. Secondly, it presents the description of the Ramsey interferometric sequence for the two-state model through the unitary evolution operator. Thirdly, it presents the theoretical description of the Gross-Pitaevskii equation (GPE) and the general background behind numerical simulations of GPE. The next section presents the development of the simulation tool which introduces the GPELabs MATLAB toolbox. Then goes on to breakdown simulated equation for the proposed interferometer along with capabilities of the developed simulation tool. Finally, the last section presents the results of the simulations.

\pagebreak
\section*{Theoretical framework}
\subsection*{Matrix description of Rabi oscillations in a two-level system}\label{Rabi model}
As first presented in \cite{Rabi1939} and carefully explained in \cite{Griffiths2005}, a collection of two-level atoms undergoes population oscillations between the two quantum states when coupled to an external oscillating electromagnetic (EM) field. These oscillations are known as Rabi oscillations and \cite{Rabi1939} provides an elegant description on how this effect is utilised for precision measurements of magnetic moments of atoms. As a brief description of the underlying process, two-level atoms create a coherent superposition of the two states evolving with time when exposed to a quasi-resonant EM field. This evolution is dependent on several major parameters such as the dipole moment of the transition, the amplitude of the EM wave and the difference (detuning) between the transition frequency and the EM frequency.

\begin{figure}[h!]
	\centering
	\includegraphics[width=0.8\linewidth]{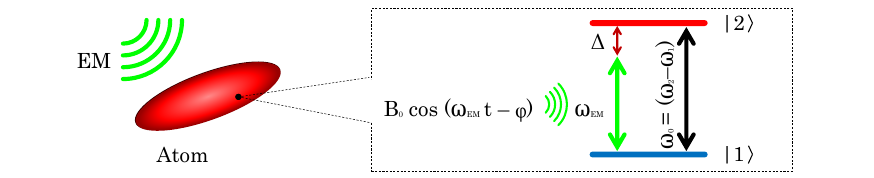}
	\caption[Schematic of two-level atom in an oscillating electromagnetic field]{Two-level atom interacting with an oscillating EM field where the energy separation between the two states is $\hbar \omega_0$.}
	\label{fig:TwoLevelAtom}
\end{figure}

Consider the two-level system of $\ket{1}$ and $\ket{2}$ quantum states shown in Figure \ref{fig:TwoLevelAtom} where the energy separation between the two states is $\hbar \omega_0$. We use this definition of the quantum states because it will be later related to the two magnetically trappable states $\ket{m_F = +1}$ and $\ket{m_F = +2}$ in the $F = 2$ hyperfine manifold of $^{87}Rb$ atoms. Consider this system being coupled to an external oscillating EM field with a frequency $\omega_{EM} = (\omega_0 - \Delta)$ where $\Delta$ is the detuning. We will consider the magnetic dipole interaction between the atom and an oscillating magnetic field, so the expression for the magnetic component of the EM field is $B_0 \cos(\omega_{EM}t-\phi) = \frac{B_0}{2} \left( e^{i(\omega_{EM}t-\phi)}+e^{-i(\omega_{EM}t-\phi)} \right)$, where $B_0$ and $\phi$ are respectively the amplitude and phase. For the state vector,
\begin{ceqn}
\begin{equation}
	\label{eqn:2LvlPsi}
	\Psi_{sys} = \sum_{i = 1}^{2} C_i \ket{i} =
	\begin{bmatrix}
	C_1\\
	C_2\\
	\end{bmatrix},
\end{equation}
\end{ceqn}
we write the time-dependent Schr\"{o}dinger equation in the rotating wave approximation
\begin{ceqn}
\begin{equation}
	\label{eqn:2LvlPsiSDE}
	i \hbar \frac{\partial \ket{\Psi_{sys}}}{\partial t} = \hat{H} \ket{\Psi_{sys}} = \left( \hat{H}_A + \hat{H}_I \right) \ket{\Psi_{sys}},
\end{equation}
\end{ceqn}
where $\hat{H}_A$ and $\hat{H}_I$ are the bare-atom and interaction Hamiltonians
\begin{ceqn}
\begin{equation}
	\label{eqn:H_Atom}
	\hat{H}_A = \frac {\hbar}{2}
	\begin{bmatrix}
		\omega_0 & 0\\
		0 & -\omega_0
	\end{bmatrix},
	\tab{}
	\hat{H}_I = \frac{\hbar}{2}
	\begin{bmatrix}
		\Delta & \Omega_R e^{-i \phi}\\
		\Omega_R e^{i \phi} & -\Delta
	\end{bmatrix},
\end{equation}
\end{ceqn}
where $\omega_0 = \omega_2-\omega_1$, $\Delta$ is the detuning of the EM field from resonance and $\Omega_R = \frac{\bra{1}\hat{\mu}\ket{2}B_0}{\hbar}$ is the Rabi frequency describing the magnetic dipole coupling of the $\ket{1}$ and $\ket{2}$ states.

We can use a matrix evolution formalism where the solution of the differential equation \ref{eqn:2LvlPsiSDE} can be replaced by a time-evolving Hamiltonian applied to $\ket{\Psi_{sys}(t)}$ at $t=0$ to obtain the time evolution of the system. Instead of using the bare state basis of Equation \ref{eqn:2LvlPsi}, we can move to the interaction picture and use the basis of the interaction Hamiltonian $\hat{H}_I$. We derive the eigenvalues and eigenvectors of $\hat{H}_I$ and apply them to the principles of unitary time evolution to obtain $\hat{H}_I(t)$. This keeps the basis states independent of time and obtains $\ket{\Psi_{sys}(t)}$ by applying the matrix operator $\hat{H}_I(t)$ to the initial conditions of the system. The eigenvalues of $\hat{H}_I$ are $\lambda_{\pm} = \pm\frac{\hbar \Omega_G}{2}$ and the eigenvectors for these values are
\begin{ceqn}
\begin{equation}
	\label{eqn:H_I_EigenVec}
	\ket{V_+} = 
		\begin{bmatrix}
		e^{\frac{-i\phi}{2}}\cos(\frac{\theta}{2})\\
		e^{\frac{i\phi}{2}}\cos(\frac{\theta}{2})
		\end{bmatrix},	
	\tab{}
		\ket{V_-} = 
		\begin{bmatrix}
		-e^{\frac{-i\phi}{2}}\sin(\frac{\theta}{2})\\
		e^{\frac{i\phi}{2}}\cos(\frac{\theta}{2})
		\end{bmatrix},
\end{equation}
\end{ceqn}
where $\tan(\theta) = \frac{\Omega_R}{\Delta}$. The application of unitary time evolution $\hat{H}_{I}(t) = e^{\frac{-i \lambda_+ t}{\hbar}} \ket{V_+} \bra{V_+} + e^{\frac{-i \lambda_- t}{\hbar}} \ket{V_-} \bra{V_-}$ leads to the unitary time evolution operator
\begin{ceqn}
\begin{equation}
	\label{eqn:H_ITime}
	\hat{U}_t = 
	\begin{bmatrix}
		\cos\left(\frac{\Omega_{G}t}{2}\right)-\frac{i\Delta}{\Omega_{G}}\sin\left(\frac{\Omega_{G}t}{2}\right) 
		& -ie^{-i\phi}\frac{\Omega_{R}}{\Omega_{G}}\sin\left(\frac{\Omega_{G}t}{2}\right)\\
		-ie^{i\phi}\frac{\Omega_{R}}{\Omega_{G}}\sin\left(\frac{\Omega_{G}t}{2}\right) 
		& \cos\left(\frac{\Omega_{G}t}{2}\right)+\frac{i\Delta}{\Omega_{G}}\sin\left(\frac{\Omega_{G}t}{2}\right) 
		\end{bmatrix}.
\end{equation}
\end{ceqn}

The evolution of the system wavefuction $\ket{\Psi_{sys}}$ when the system starts in state $\ket{2}$ (e.g., the $\ket{m_F = +2}$ state of the $^{87}Rb$ atom) can be obtained by applying the $\hat{U}_t$ operator to the initial state vector $\begin{bmatrix} 0 & 1 \end{bmatrix}^T$. This leads to
\begin{ceqn}
\begin{equation}
	\label{eqn:Psi_Sys2}
	\ket{\Psi_{sys}(t)} = \hat{U}_t \begin{bmatrix} 0\\ 1 \end{bmatrix} = 
	\begin{bmatrix}
	-ie^{-i\phi}\frac{\Omega_{R}}{\Omega_{G}}\sin\left(\frac{\Omega_{G}t}{2}\right)\\
	\cos\left(\frac{\Omega_{G}t}{2}\right)+\frac{i\Delta}{\Omega_{G}}\sin\left(\frac{\Omega_{G}t}{2}\right) 
	\end{bmatrix}.		
\end{equation}
\end{ceqn}

\subsection*{Ramsey method of separated oscillatory fields}\label{RamseyIntro}
Following the work of I.I. Rabi, N.F. Ramsey \cite{Ramsey1990} significantly improved the Rabi method by using two oscillatory fields with a short pulse length $\tau$ separated by a long free evolution time $T$ to study molecular resonances and demonstrated linewidths that are \num{0.6} times narrower compared to that of the single oscillatory field method. This work of N.F. Ramsey won him the Nobel prize in physics in \num{1989} and is commonly referred to as Ramsey interferometry in the physics community. The Ramsey interferometric method provides the basis of the exquisite time standards using $Cs$ fountain clocks. As examples, the NIM5 $Cs$ fountain clock \cite{Fang2015} based in China has an uncertainity of \num{1.6e-15} and the NIST-F1 $Cs$ fountain clock \cite{Heavner2005} based in the USA has an uncertainty of \num{0.97e-15}. These uncerainities translate to a \SI{1}{\s} error in time keeping for every \num{20} to \num{30} million years. Further, Ramsey interferometry is utilised to obtain extremely sensitive measurements of local gravity as reported in \cite{Peters1999} and a Ramsey-type method with a spin-echo pulse (i.e. a $\pi$-pulse during the free evolution time $T$) is utilised in \cite{Rosi2014} to obtain precise measurements of the Newtonian gravitational constant $G = \SI{6.67191(99)e-11}{\m^3 \kg^{-1} \s^{-2}}$.

A typical  Ramsey sequence interrogates a single two-level atom with a single-frequency quasi-resonant electromagnetic field. Briefly, Ramsey interferometry synchronizes two independent oscillators (an electromagnetic field and a coherent superposition of two quantum states of an atom) and compares their evolution in time. The atom is initially in state $\ket{2}$. The first $\frac{\pi}{2}$-pulse of duration $\tau$ (Rabi method) prepares a coherent superposition of the two states with equal (\num{50}-\num{50}) populations which is synchronized with the EM radiation. Then the atom-EM coupling is turned off and the two oscillators evolve independently for the evolution time $T$. The second (recombination) $\frac{\pi}{2}$-pulse interrogates the superposition state and extracts the phase difference (the relative phase) between the two quantum states. This relative phase is contained in the population difference of the states \cite{Anderson2009}. Further, the dynamical behavior of the two-level system wavefunction and its behaviour between the two states can be mathematically represented via the Bloch vector formalism \cite{Feynman1957}. 

As derived in Equation \ref{eqn:H_ITime}, the unitary time operator $\hat{U}$ can be used to determine the system wavefunction $\Psi_{sys}(t)$ at the end of each step of the Ramsey interferometric sequence. One of the unitary time evolution operators relevant to the sequence is the operation for the combination and the recombination pulses of the superposition. A $\num{50}:\num{50}$ superposition in a two-level system is created when $\Omega_{G}t=\frac{\pi}{2}$. Applying this condition to $\hat{U}_t$ in Equation \ref{eqn:H_ITime} leads to the following expression for the unitary evolution operator $\hat{U}_{\tau}$ for the $\frac{\pi}{2}$-pulse. The other important case is the free evolution operator $\hat{U}_T$ which is when $\Omega_R = 0$ for $\hat{U}_t$ in Equation \ref{eqn:H_ITime} leading to
\begin{ceqn}
\begin{equation}
	\label{eqn:U_pi_T}
	\hat{U}_{\tau} = \frac{1}{\sqrt{2}}
	\begin{bmatrix}
	1 & -ie^{-i\phi}\\
	-ie^{i\phi} & 1
	\end{bmatrix}, 
	\tab{}
	\hat{U}_{T} =
	\begin{bmatrix}
	e^{-\frac{i\Delta T}{2}} & 0\\
	0 & e^{\frac{i\Delta T}{2}}
	\end{bmatrix}.
\end{equation}
\end{ceqn}
The evolution operators above are sequentially applied to the initial state of $\Psi_{sys}(0)$ as 
\begin{ceqn}
\begin{equation}
	\Psi_{sys}(t) = \hat{U}_{\tau}.\hat{U}_{T}.\hat{U}_{\tau}.\Psi_{sys}(0),
\end{equation}
\end{ceqn}
to obtain an expression for $\Psi_{sys}(t)$ at the end of the Ramsey sequence. This leads to
\begin{ceqn}
\begin{equation}
	\label{eqn:Psi_t}
	\Psi(\tau+T+\tau)=
	\begin{bmatrix}
	-ie^{\frac{-i\phi}{2}}\cos(\frac{\Delta T-\phi}{2})\\
	ie^{\frac{i\phi}{2}}\sin(\frac{\Delta T-\phi}{2})
	\end{bmatrix},
\end{equation}
\end{ceqn}
where $T$ is the free evolution time, $\Delta$ and  $\phi$ are respectively the detuning of the EM field from resonance and the phase of the second $\frac{\pi}{2}$-pulse.

\begin{figure}[h!]
	\centering
	\includegraphics[width=0.8\linewidth]{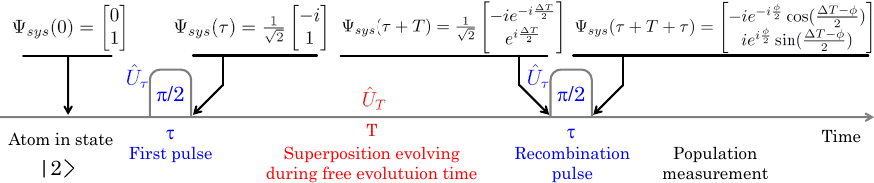}
	\caption[Evolution of $\Psi_{sys}(t)$ of a two-level system during Ramsey interferometry]{Evolution of the state vector $\Psi_{sys}(t)$ of a two-level atom in a typical Ramsey interferometric sequence.}
	\label{fig:RamseyEqn}
\end{figure}

The intermediate expressions for $\Psi_{sys}(t)$ at different stages of the Ramsey sequence are shown in Figure \ref{fig:RamseyEqn}. The equation for $\Psi_{sys}(t)$ at the end of the Ramsey sequence depends on the free evolution time $T$, the detuning of the EM field from resonance $\Delta$ and the phase $\phi$ of the second $\frac{\pi}{2}$-pulse. These are the three domains (the time domain, the frequency domain and the phase domain) in which Ramsey interferometry can be preformed. It should be noted that Ramsey interferometry in the frequency domain may be riddled by the shift of the detuning over the course of the experiment and frequency instability (e.g., magnetic noise) in the equipment. However, the scheme proposed in \cite{Vitanov2015} is an elegant method to utilise Ramsey interferometry in the frequency domain. Further, Ramsey interferometry is also performed in the atom number $N$ domain \cite{Egorov2011}. Here, we utilises Ramsey interferometry in the phase domain where $\phi$ of the second $\frac{\pi}{2}$-pulse is manipulated to extract the interferometric data between $\ket{1}$ and $\ket{2}$ for a fixed $T$. The fixed $T$ is one of the major advantages of operating in the phase domain as it avoids the shot-to-shot variation of atom numbers due to trap losses for varying hold times.

The interesting aspect in this work is the state dependent spatial separation of the two states due to gravitational sag based on Equation \ref{eqn:G_Sag}. This causes the superposition of states to spatially separate and recombine during $T$. The overlap of the two components before the recombination pulse defines the visibility of the Ramsey fringe as denoted by Equation \ref{eqn:RamseyCon} \cite{Egorov2011}. At the end of the Ramsey sequence we introduce a variable $P_z(T,\Delta,\phi)$ where,
\begin{ceqn}
\begin{equation}
	\label{eqn:RamseyCon}
	P_z(T,\Delta,\phi) = \frac{2}{N} Im \left[ e^{i(\phi+\Delta T)} \int \Psi_{(\tau + T)} ^{\ket{+2}*} \Psi_{(\tau + T)} ^{\ket{+1}} \right] d^3r,
\end{equation}
\end{ceqn}
which contains all dependencies and can be measured via an absorption image.

Next we shall investigate the theoretical framework that describes interactions of the proposed two-level Ramsey interferometric sequence.

\subsection*{The Gross-Pitaevskii equation and its numerical simulation}\label{GPEForm}
The dynamics of the $\ket{+1}-\ket{+2}$ BEC system during the Ramsey interferometric sequence is analysed via the Gross-Pitaevskii equation (GPE). This foundational theoretical model was introduced by E. P. Gross and L. P. Pitaevskii in 1961 \cite{Gross1961} for analysing the dynamical evolution of the zero temperature condensate wavefunction. The GPE is an extension of the standard Schr\"{o}dinger equation which incorporates interactions between atoms in a dilute cloud through a mean-field interaction potential $U_{int}(r)$. The GPE model has well-defined many experimental data whilst reducing the calculation power necessary as shown in Figure 4 in \cite{Anderson2009}. The GPE is also used complementarily with experimental results to extract enhanced measurements as shown in \cite{Egorov2013,Egorov2011}.

The time-dependent GPE takes the form of Equation \ref{eqn:GPE_TD} \cite{Pitaevskii2003}, which describes both the temporal and spatial dynamics of the BEC system.
\begin{ceqn}
\begin{equation}
	\label{eqn:GPE_TD}
	i \hbar \frac{\partial \Psi(r,t)}{\partial t} = - \frac{\hbar^2}{2m}\triangledown^2\Psi(r,t)+V_{HO}(r) \Psi(r,t) + U_0 |\Psi(r,t)|^2 \Psi(r,t),
\end{equation}
\end{ceqn} 
where $\Psi(r,t)$ is the time-dependent wavefunction, $\triangledown^2 = \frac{\partial^2}{\partial r^2}$ is Laplace operator, $V_{HO}$ is the external harmonic oscillator trapping potential and $U_0 $ is the effective interaction parameter.

The first term in the equation encapsulates the kinetic energy of the system and the second term captures the potential energy via the external harmonic potential defined by $ V_{HO} = \frac{m}{2}\left(2\omega_r^2 r^2 + \omega_y^2y^2 \right)$ for a cigar-shaped trap, where $\omega_{r,y}$ are the radial and axial trap frequencies, respectively. The last term in the equation describes the energy variation due to interactions with the special parameter $U_0$ know as the interaction constant defined as $U_0 = \frac{4\pi\hbar^2a_s}{m}$, where $a_s$ is the s-wave scattering length between the interacting atoms.

The GPE in the form of Equation \ref{eqn:GPE_TD} describes the temporal and spatial evolution of a single BEC with internal collisional interactions. For the two-component $\ket{+1}-\ket{+2}$ BEC system, the GPE requires to include intra-state (within the states) and inter-state (between states) interactions resulting in a set of equations that describes a two-component BEC. The intra-state interactions are described by $U_{ii} = \frac{4 \pi \hbar^2 a_{ii}}{m}$, where $a_{ii}$ is the intra-state scattering length for elastic collisions between atoms of the same state $i$. The inter-state interactions are described by $U_{ij} = \frac{4 \pi \hbar^2 a_{ij}}{m}$ , where $a_{ij}$ is the inter-state scattering length for elastic collisions between atoms of adjacent states $ij$. This lead to coupled equations for a two-component interacting BEC system of the form 
\begin{ceqn}
\begin{equation}
	\label{eqn:GPE_2CBEC}
	\begin{split}
	i \hbar \frac{\partial \Psi_1(r,t)}{\partial t} &= - \frac{\hbar^2}{2m}\triangledown^2\Psi_1(r,t)+ V_{HO 1}(r) \Psi_1(r,t) + U_{11} |\Psi_1(r,t)|^2 \Psi_1(r,t) + U_{12} |\Psi_2(r,t)|^2 \Psi_1(r,t),\\
	i \hbar \frac{\partial \Psi_2(r,t)}{\partial t} &= - \frac{\hbar^2}{2m}\triangledown^2\Psi_2(r,t) + V_{HO 2}(r) \Psi_2(r,t) + U_{22} |\Psi_2(r,t)|^2 \Psi_2(r,t) + U_{12} |\Psi_1(r,t)|^2 \Psi_2(r,t),
	\end{split}
\end{equation}
\end{ceqn}
which can be solved via numerical methods. Further, each equation has parameters unique to either state allowing to incorporate state-dependent external trapping potentials, where effects of gravitational sag are incorporated into the analysis. Also, additional aspects of interactions with an external EM field and atom losses from either state will be added leading to an expanded analysis.

Numerical simulations are of great importance to obtain an insight into the $\ket{+1} - \ket{+2}$ BEC dynamics. As a precursor, the coupled Gross-Pitaevskii equations in Equation \ref{eqn:GPE_2CBEC} are useful to analyse the impact of the inter-state interactions on the suggested Ramsey sequence as well as to probe optimal conditions to maximise the Ramsey contrast at the end of one dipole oscillation. Though the in-house developed xSPDE Matlab simulation toolbox is capable of simulating these coupled partial differential GPEs  \cite{Kiesewetter2016}, the theoretical framework presented in \cite{Bao2003_1,Bao2003_2} is used as the cornerstone for the numerical simulations in this project. The main advantage of \cite{Bao2003_1,Bao2003_2} is the readily available GPELabs MATLAB tool box \cite{Antoine2014, Antoine2015} which does not require developing the simulation environment compared with xSPDE. Further, xSPDE is tailored towards studying stochastic effects which is not considered here.

The GPELabs MATLAB tool box utilises several finite difference and spectral methods to calculate the ground-state and the dynamics of wavefunctions for multi-component Bose-Einstein condensate systems. To calculate the ground-state, GPELabs offers four schemes: backward Euler finite difference scheme (BEFD), Crank-Nicolson finite difference scheme (CNFD), backward Euler spectral discretization scheme (BESP) and Crank-Nicolson spectral discretization scheme (CNSP). From these the first two are finite difference schemes and the second two are spectral schemes. To calculate the dynamics, GPELabs has six schemes with two additional pseudo spectral schemes to the above which are time splitting pseudo spectral scheme (TSSP) and relaxation pseudo spectral scheme (ReSP).

An overall summary of the numerical schemes relating to solving BEC systems is presented in \cite{Minguzzi2004}. It reports a vast variety of schemes spanning the areas of simulating BECs at zero temperature, BECs at finite temperatures with a thermal fraction, as well as systems of fermions and boson-fermion mixtures. Here we are focused on a pure BEC system without any thermal fraction. A detailed study of the TSSP method is presented in \cite{Bao2002} and an extensive descriptions of both BEFD and TSSP are presented in \cite{Bao2004_1} which reserves the BEFD for ground-state calculations with proven energy diminishing condition under normalized gradient flow and reserves the TSSP for dynamical simulations of BECs. The simulated ground-states via BEFD are confirmed by comparing with the experimental results for the two component BEC system in the immiscible regime in \cite{Hall1998_1} and goes on to simulate the dynamics of the same system via TSSP showing the separation and recombination of the two components due to miscibility. In \cite{Bao2004_2} the numerical results from BEFD and TSSP are compared with CNFD and forward Euler finite difference (FEFD) methods. It shows that BEFD and TSSP are much better in preserving the energy diminishing property under continuous normalized gradient flow and reports the importance of picking the appropriate spatial mesh and time-step size to obtain correct results via CNFD and FEFD schemes. The work presented in \cite{Antoine2013, Bao2013, Bao2014} outlines all the calculation methods discussed above in several avenues; however, the optimal method for ground-state and dynamic calculations of coupled GPEs is considered. Regarding ground-state calculations, \cite{Bao2013, Bao2014} compares the BEFD and BESP methods and concludes BESP is better as it requires fewer grid points which saves computational memory and time especially for the 3D scenario. Regarding the dynamic scenario, the framework for the ReSP scheme was first introuduced in \cite{Besse2004}, which reports a vast reduction in computational time for the relaxation scheme compared to that of CNFD and highlights its ability to conserve the density and energy compared to the TSSP method. For these reasons, the BESP scheme is selected to calculate the ground-state and the ReSP scheme is selected to calculate the dynamics of the two component coupled GPEs. Further, scheme dependent variations of the simulated outputs are considered out of the current scope.

\section*{The development of the simulation tool}\label{GPELabs}
MATLAB was used as the primary software tool to simulate the two coupled Gross-Pitaevskii equations for the two component BEC system due to the dedicated GEPLab toolbox developed by Xavier Antoine and Romani Duboscq \cite{Antoine2014, Antoine2015}.  This toolbox has the capability to solve the GPE to find the wavefunction of the ground-state and the time-dependent behaviour (including Rabi coupling and dipole oscillations) for a single or multi component BEC system in 1D, 2D or 3D. 

The GPE is calculated in the dimensionless form in GPELabs which vastly expands its applicability for BEC systems with non-linear Schr\"{o}dinger equations. The dimensionless form of the GPE as per \cite{Antoine2014} is
\begin{ceqn}
\begin{equation}
	\label{eqn:GPE_DimLss}
	\left\lbrace
	i \frac{\partial\Psi(r,t)}{\partial t} = \left(\frac{-1}{2}\Delta + V + \beta|\Psi(r,t)|^2 - \Omega L_z
	\right) \Psi(r,t) \right\rbrace,
\end{equation}
\end{ceqn}
where mathematical expressions within curly braces ($\{Expression\}$) or variables with a bar ($\bar{X}$) denotes the dimensionless form.

An important facet of the simulation tool is the conversion between the dimensionless values and the real values. It is important to encode the experimental scenario into the simulation and then to decode simulated results into real values. The experimental trap frequencies provide the footing for the unit of time for the dimensionless framework where $\omega_m = \min(\omega_{x,y,z})$. The unit length is then defined via the harmonic oscillator characteristic length $a_0 = \sqrt{\frac{\hbar}{m \omega_m}}$, \cite{Antoine2014}. These allow to define the dimensionless unit of time by $\bar{t} = \omega_m t$ and the dimensionless unit of length $\bar{r}=\frac{r}{a_0}$. Based on these the 3D wavefunction has the conversion $\bar{\Psi}_{3D}(\bar{r},\bar{t}) = a_0^\frac{3}{2}\Psi_{3D}(r,t)$. It should be noted that the unit of $\Psi(r,t)$ is dependent upon the dimensionality such that $[\Psi_{nD}] = [L]^{\frac{-n}{2}}$. Therefore, care must be taken when converting the GPE between dimensions. All frequencies are relative to $\omega_m$ where $\bar{\Omega}=\frac{\Omega}{\omega_m}$. Finally, the dimensionless energy takes the form as $\bar{E} = \frac{E}{\hbar \omega_m}$. These relations are of great importance to correlate the experimental scenario with simulated results. Table \ref{tbl:GPELab_DltoReal} provides a summary of the conversion factors that convert simulated results to real values where the dimensionless values are denoted by $\bar{X}$.
\begin{table}[h!]
	\centering
	\begin{tabular}{|c|c|} 
 	\hline
 	Real value & Converting form simulations \\ [0.5ex] 
 	\hline
	$t$	& $\frac{\bar{t}}{\omega_m }$ \rule{0pt}{2.6ex} \\ [0.5ex] 
 	\hline
	$r$ & $a_0 \bar{r}$ \rule{0pt}{2.6ex} \\ [0.5ex] 
	\hline
	$\Psi_{3D}(r,t)$& $a_0^\frac{-3}{2} \bar{\Psi}_{3D}(\bar{r},\bar{t})$ \rule{0pt}{2.6ex} \\ [0.5ex] 
 	\hline
 	$\Omega$ & $\bar{\Omega} \omega_m$ \rule{0pt}{2.6ex} \\ [0.5ex] 
 	\hline
 	$E$ & $\bar{E}\hbar \omega_m$ \rule{0pt}{2.6ex} \\ [0.5ex] 
 	\hline
	\end{tabular}
\caption{Conversions of parameters from simulation to real values where the dimensionless values from the simulation are denoted by $\bar{X}$}
\label{tbl:GPELab_DltoReal}
\end{table}

Furthermore, the simulated system of wavefunctions is normalised to unity within GPELabs where $\int \Psi_{sys}^* \Psi_{sys} = 1$. However, most of the general literature normalise a system of wavefunctions to the total population $N$ where $\int \Psi_{sys}^* \Psi_{sys} = N$. This should be carefully considered when adapting terms into GPELabs as some terms acquire a factor of $N$ to the standard expression. Due to the complicated nature of converting a 3D experimental to a 1D approximation, the simulation was developed for the 3D scenarios. 

The terms relevant to this work in Equation \ref{eqn:GPE_DimLss} are the Laplace term or the dispersion retaliation ($\bar{\Delta}$), the linear term ($\bar{V}$) and  the non-linear term ($\{\beta|\Psi(r,t)|^2\}$). The angular momentum term for the BEC ($\{\Omega L_z\}$) is not considered, leading an equation with the following structure
\begin{ceqn}
\begin{equation}
	\label{eqn:GPE_Fromat}
	\left\lbrace
	\begin{bmatrix}
	i\dot{\Psi}_1\\
	i\dot{\Psi}_2
	\end{bmatrix}= 
	\left(
	\frac{-1}{2}
	\begin{bmatrix}
	\Delta_{11} & 0\\
	0 & \Delta_{22}
	\end{bmatrix} 
	+
	\begin{bmatrix}
	L_{11} & L_{12}\\
	L_{21} & L_{22}
	\end{bmatrix}
	 +
	\begin{bmatrix}
	NL_{11} & NL_{12}\\
	NL_{21} & NL_{22}
	\end{bmatrix}
	\right)
	\begin{bmatrix}
	\Psi_1\\
	\Psi_2
	\end{bmatrix}
	\right\rbrace,
\end{equation}
\end{ceqn}
where $\bar{\Psi}_n$ is the dimensionless form of the wavefunction for state $\ket{+n}$, $\bar{\Delta}_{ii}$ are dispersion relations, $\bar{L}_{ij}$ are the linear terms and $\bar{NL_{ij}}$ are non-linear terms. The advantage of this interpretation is the ease with which the linear terms and the non-linear terms can be compartmentalised to simulate the BEC system.

The conversion of dimensionless parameters is well described in \cite{Bao2003_1} which shows the adaptation of equations in 1D, 2D and 3D. The work in \cite{Bao2003_1} shows a dependency of selecting a characteristic length scale with the interaction strength of the simulated scenario. In short, $a_0$ is a valid selection if the BEC is in the weak interaction regime where $4\pi|a_s|N \ll a_0$ or in the moderate interaction regime where $4\pi|a_s|N \approx a_0$, where $a_s$ is the relevant scattering length and $N$ is the atom number. If the simulation is in the strong interaction regime where $4\pi|a_s|N \gg a_0$, a better characteristic length $x_s$ is suggested instead of $a_0$ where $x_s = (4 \pi|a_s|N a^4_0 )^\frac{1}{5}$ and goes on to explain that the selection of the appropriate characteristic length depicts the visibility of phenomena and what can be resolved by the discretization of specified spatial and temporal grids. This does not appear in the GPELabs documentation; however, GPELabs provides full control over this discretization over space and time which leads to needing several trails to set-up the appropriate grid to observe the dynamics of the system. Based on the conversion factors in Table \ref{tbl:GPELab_DltoReal}, the simulated scenario can be extracted out to the real world.

The outputs of GPELabs are in both numerical and graphical forms for all components. The numerical outputs are the square of the wavefunction at the origin $|\bar{\Psi}_{3D}(0,\bar{t})|^2$, the directional rms-radii, the energy and the chemical potential of each component. The graphical outputs are the three-dimensional distribution of the square of the wavefunction $|\bar{\Psi}_{3D}(\bar{r},\bar{t})|^2$ and the phase for each component. The frequency of acquiring this information can be set at desired intervals of iterations. 

As suggested above, the separation of linear terms and non-linear terms can be done when all simulated phenomena are combined and adapted to the framework of GPELabs. The linear terms $\bar{L}$ take the form 
\begin{ceqn}
\begin{equation}
	\label{eqn:GPE_Lin}
	\left\lbrace
	\begin{bmatrix}
	L_{11} & L_{12}\\
	L_{21} & L_{22}
	\end{bmatrix}
	= 
	\begin{bmatrix}
	V^{HO}_{11} + V^{EM}_{11} & V^{EM}_{12}\\
	V^{EM}_{21} & V^{HO}_{22} + V^{EM}_{22}
	\end{bmatrix}
	\right\rbrace,
\end{equation}
\end{ceqn}
which accounts for the state dependent harmonic oscillator (HO) trapping potential ($\bar{V}^{HO}_{i,i}$) and the interaction with the external EM field ($\bar{V}^{EM}_{ij}$) as both of these are multiplied linearly by the wavefunction. The HO potential takes up the diagonal terms where the interaction with the EM field takes up both diagonal and off-diagonal terms. When the EM field is applied, the off-diagonal elements couple the two states of the wavefunction into a superposition of states. The non-linear terms $\bar{NL}$ take the form
\begin{ceqn}
\begin{equation}
	\label{eqn:GPE_NonLin}
	\left\lbrace
	\begin{bmatrix}
	NL_{11} & NL_{12}\\
	NL_{21} & NL_{22}
	\end{bmatrix}
	= 
	\begin{bmatrix}
	V^{INT}_{11} + V^{Loss}_{11} & 0\\
	0 & V^{INT}_{22} + V^{Loss}_{22}
	\end{bmatrix}
	\right\rbrace,
\end{equation}
\end{ceqn}
where the off-diagonal elements are zero and the diagonal terms account for both interactions ($\bar{V}^{INT}_{i,i}$: inter- and intra-state) and loss ($\bar{V}^{Loss}_{i,i}$: two- and three-body) terms as both of these are multiplied by a higher-order of the wavefunction. The inclusion of the loss terms expand the capabilities of the simulation tool, but are not considered in the primary results. Within the considered theoretical framework, the non-linear interactions are impacted by the populations of either state. Further, higher-order couplings that lead to effects such as the Josephson effect \cite{Tacla2010} are not considered. 

\subsection*{Adapting linear terms to the GPELabs framework}
The linear terms of the simulated phenomena are the external HO trapping potential with gravitational sagging and the interactions with the EM fields. The external trapping potential has the form $V^{HO} = \frac{m}{2} \left( \omega^2_x x^2 + \omega^2_y y^2 +\omega^2_z z^2  \right)$. Further, when the effects of gravitational sagging based on Equation \ref{eqn:G_Sag} are added, $V^{HO} = \frac{m}{2} \left( \omega^2_x x^2 + \omega^2_y y^2 +\omega^2_z (z-\Delta z_{sag})^2  \right)$. The dimensionless form of $\bar{V}^{HO}$ is 
\begin{ceqn}
\begin{equation}
\resizebox{0.9 \textwidth}{!} 
{$
	\label{eqn:GPE_V_HO_Mat}
	\bar{V}^{HO}
	 = \left\lbrace
	\begin{bmatrix}
	\frac{1}{2} \left( \gamma^2_{1x} x^2 + \gamma^2_{1y} y^2 +\gamma^2_{1z} (z-\Delta z_{1sag})^2  \right) & 0\\
	0 & \frac{1}{2} \left( \gamma^2_{2x} x^2 + \gamma^2_{2y} y^2 +\gamma^2_{2z} (z-\Delta z_{2sag})^2  \right)
	\end{bmatrix}
	\right\rbrace,
$}
\end{equation}
\end{ceqn}
where $\bar{\gamma}_{x,y,z} = \frac{\omega_{x,y,z}}{\omega_m}$ are dimensionless directional trap frequencies and $\Delta \bar{z}_{sag} = \frac{\Delta z_{sag}}{a_0}$ is the dimensionless gravitational sag of each component

\begin{figure}[h!]
	\centering
	\includegraphics[width=0.8\linewidth]{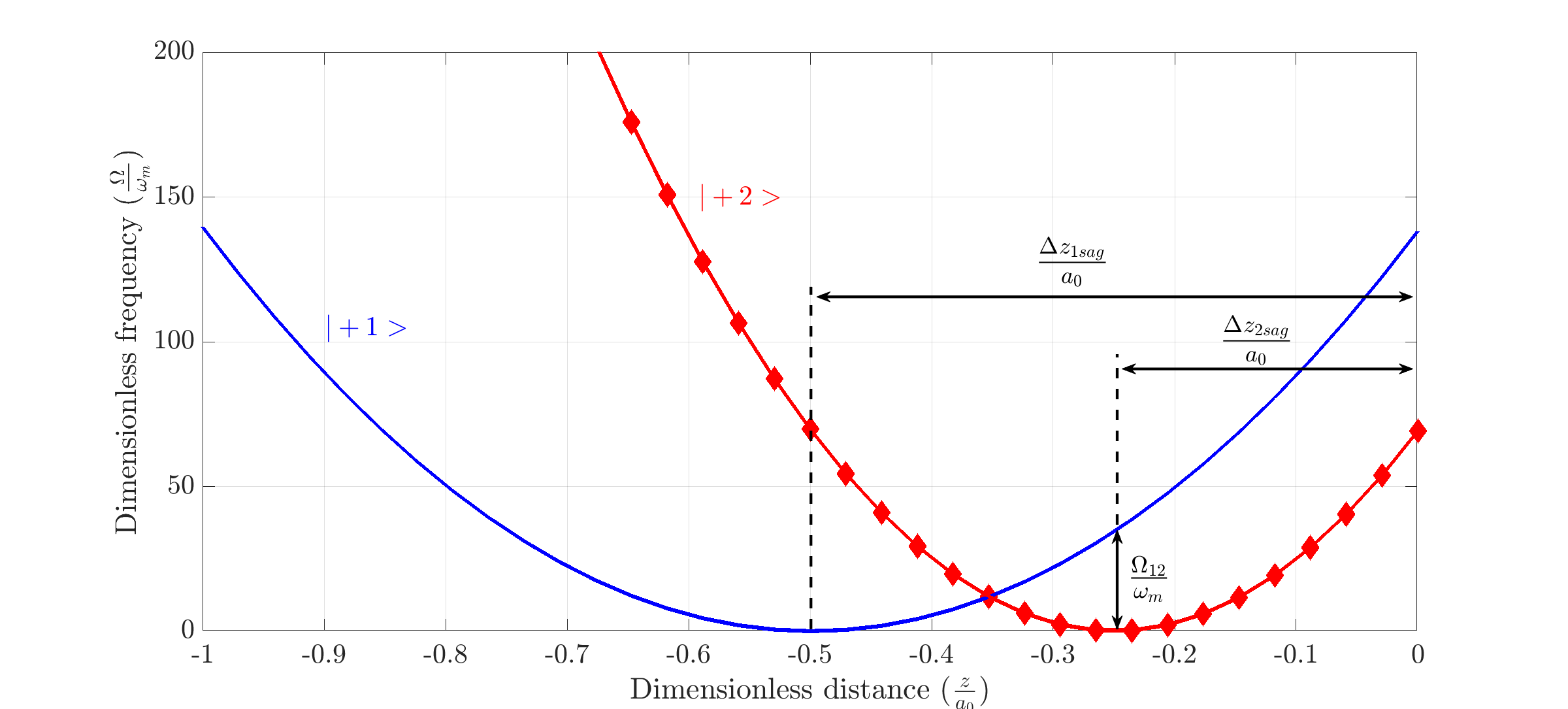}
	\caption[GPELabs harmonic oscillator potential]{External vertical trapping potential with gravitational sagging experienced by $\ket{+1}$ (blue) and $\ket{+2}$ (red) with trap frequency $\omega_r = 2 \pi \times \SI{400}{\Hz}$ for the $\ket{+1}$ state.}
	\label{fig:V_HO_Gsag}
\end{figure}

A simulation for a harmonic oscillator potential with $\omega_r = 2 \pi \times \SI{400}{\Hz}$ for state $\ket{+1}$ in blue and state $\ket{+2}$ in red is shown in the Figure \ref{fig:V_HO_Gsag}. The horizontal axis of the graph is the dimensionless vertical distance $z/a_0$ from the atom chip. The $y$-axis of the graph is energy in dimensionless radial frequency units ($\bar{\Omega}$). A key feature displayed in the figure is the spatial separation of the trapping potentials due to gravitational sagging. Moreover, an important insight is the energy gap $\bar{\Omega}_{12}$  felt by atoms in state $\ket{+1}$ at its starting point which is the potential minimum $\Delta z_{2sag}$ of state $\ket{+2}$. This energy gap should be accounted for when applying the EM field as otherwise the atoms in state $\ket{+1}$ are untrapped hindering the population transfer in creating a superposition.

The other linear term that influences this project is the interaction term with the external EM field. This follows Equation \ref{eqn:H_Atom} leading to $V^{EM} = \hat{H}_I$. The dimensionless form for this can also be derived by converting the standard GPE into the dimensionless form
\begin{ceqn}
\begin{equation}
	\label{eqn:V_EM}
	\bar{V}^{EM}
	=
	\frac{1}{2\omega_m}
	\begin{bmatrix}
	0 & \Omega_R e^{-i \phi}\\
	\Omega_R e^{i \phi} &	-2\Delta 
	\end{bmatrix}
	=	
	\frac{1}{2}
	\left\lbrace
	\begin{bmatrix} 
	0 & \bar{\Omega}_R e^{-i \phi}\\
	\bar{\Omega}_R e^{i \phi} &	-2\bar{\Delta}
	\end{bmatrix}
	\right\rbrace,
\end{equation}
\end{ceqn}
where $\bar{\Delta} = \frac{\Delta}{\omega_m}$ is the dimensionless detuning of EM field from the resonance between $\ket{+1}$ and $\ket{+2}$ and $\bar{\Omega}_R = \frac{\Omega_R}{\omega_m}$ is the dimensionless Rabi frequency for RF coupling between the two states. 

An interesting point is the lack of a conversion factor for the relative phase $\phi$ of the EM field when converting to its dimensionless form. Therefore, the value of $\phi$ in the simulation is the value for the phase in the real experiment. There is a subtle connection between the two linear terms: the trapping potential and the applied EM field. As per Figure \ref{fig:V_HO_Gsag} gravitational sagging causes a separation of the two potential minima for either state of the $\ket{+1}-\ket{+2}$ system. This leads to an energy gap between the two states at $\Delta\bar{z}_{2sag}$ where the $\ket{+2}$ BEC is created. This gap needs to be filled to bring the two states on-resonance prior to applying the mixing pulse via the EM field. If the EM field is applied off-resonance, then the transferred atoms from $\ket{+2} \rightarrow \ket{+1}$ are immediately lost as there is no trap at $\Delta \bar{z}_{2sag}$ for $\ket{+1}$ atoms. The detuning $\bar{\Delta}$ of the EM field is utilized to bring the two states to on-resonance, where $\bar{\Omega}_{12} = 2\bar{\Delta}$ at which point the transfer of atoms from $\ket{+2} \rightarrow \ket{+1}$ is most efficient. The detuning $\Delta$ required for the EM field of the real experiment can be extrapolated out from $\bar{\Omega}_{12}$ in the simulation as $\bar{\Delta} = \frac{\bar{\Omega}_{12}}{2}$ and $\Delta = \frac{\bar{\Omega}_{12}}{2}\omega_m$, where $\bar{\Omega}_{12}$ is readily available via the simulation. This may be of great assistance to provide an estimated starting point for the experiment.

\subsection*{Adapting non-linear terms to the GPELabs framework}
There are two non-linear terms incorporated into the simulation tool which are the interaction term $V^{INT}_{ii}$ and the loss term $V^{Loss}_{ii}$. The general form of the interaction term follows $V^{INT} = U_{ij}|\Psi_k|^2$, where $U_{ij} = \frac{4\pi\hbar^2 a_{ij}}{m}$ and $i,j,k \in \{1,2\}$. The dimensionless form of $\bar{V}^{INT}$ when $\int \Psi_{sys}^* \Psi_{sys} = 1$ is
\begin{ceqn}
\begin{equation}
	\label{eqn:GPE_V_INT}
	\bar{V}^{INT}
	 = 
 	\left\lbrace
 	4\pi N
	\begin{bmatrix}
	a_{11} |\Psi_1|^2 + a_{12} |\Psi_2|^2 & 0\\
	0 &	a_{22} |\Psi_2|^2 + a_{12} |\Psi_1|^2
	\end{bmatrix}
	\right\rbrace,
\end{equation}
\end{ceqn}
where $a_{ij}$ are the scattering lengths between the interacting two states, $\bar{a}_{ij} = \frac{a_{ij}}{a_0}$ and $N$ is the total atom number of the $\ket{+1}-\ket{+2}$ system.

Further, careful consideration should be given to $\bar{V}^{INT}$ when converting to 1D and 2D simulations. As described on pages 324 - 339 of \cite{Pitaevskii2003}, there is a dimensional dependence on the interaction constant $U_{ij}$. For the 1D case, $\bar{U}^{1D} = \frac{2 \hbar^2}{m} \frac{a_s}{a^2_\perp}$, where $a_\perp = \sqrt{\frac{\hbar}{m \omega_\perp}}$. This is to be combined with the dimensionality of the wavefunction when converting $U^{1D}$ into the dimensionless form where $\bar{\Psi}^{1D} = a_\perp^\frac{1}{2}\Psi^{1D}$. Further, for the 2D case,  $\bar{U}^{2D} = \frac{\sqrt{8\pi} \hbar^2}{m} \frac{a_s}{a_z}$, where $a_z = \sqrt{\frac{\hbar}{m \omega_z}}$. Also, similar to the 1D case, the dimensionality of the wavefunction should be accounted for when converting $U^{2D}$ into the dimensionless form where $\bar{\Psi}^{2D} = a_\perp\Psi^{2D}$.

Moreover, the additional aspect of loss terms $V^{Loss}_{ii}$ was coded into the simulation, where the two-body and three-body loss terms take the general form of $V^{Loss} = -i \hbar \Gamma$. By compacting all the loss terms together and converting the 3D GPE to its dimensionless form along with the normalisation  $\int \Psi_{sys}^* \Psi_{sys} = 1$ results in
\begin{ceqn}
\begin{equation}
	\label{eqn:V_Loss_2}
	\bar{V}^{Loss}_{ii}
	= 
	\begin{bmatrix}
	\frac{-iN(\xi_{11} |\Psi_1|^2+ \xi_{12} |\Psi_2|^2)}{2 \omega_m a^3_0} & 0\\
	0 & \frac{-iN \xi_{12} |\Psi_1|^2}{2 \omega_m a^3_0} + \frac{-iN \xi_{222} |\Psi_2|^4}{2 \omega_m a^6_0},
	\end{bmatrix},
\end{equation}
\end{ceqn}
where $N$ is the total atom number, $\xi$ are two- and three-body loss terms, $\omega_m$ is the minimum trap frequency and $a_0$ is the harmonic oscillator characteristic length. The two-body loss term for $\ket{+2}$ with $\xi_{12} = 0$ for the $\ket{+1}-\ket{+2}$ interaction as this interaction can only switch the states between the two atoms causing no loss. The two-body loss rate for $\ket{+1}$ with loss coefficient $\xi_{11} = \SI{8.1(3)e-14}{\cm^3 \per \s}$ \cite{Egorov2013} for the $\ket{+1}-\ket{+1}$ interactions. The three-body loss rate for $\ket{+2}$ with the loss coefficient $\xi_{222} = \SI{1.8e-29}{\cm^6 \per \s}$ \cite{Sding1999} which is the only loss channel via three-body recombination of three $\ket{+2}$ atoms.

\section*{Simulations of Ramsey interferometry in the trapped $\ket{+1}$ - $\ket{+2}$ BEC system}\label{RamseySimulations}

The BEC in the experiment has a total atom number of about $N = \SI{3e4}{}$ and the conditions were selected such that estimated axial and radial trapping frequencies for state $\ket{+1}$ are $\omega_{ax} \approx 2 \pi \times \SI{400}{\Hz}$ and $\omega_{r} \approx 2 \pi \times \SI{12}{\Hz}$. Here, the notations for the properties of states $\ket{+1}$ and $\ket{+2}$ are represented by subscripts $X_1, X_2$. The values for the scattering length were set as $a_{11} =95.44 a_0$ and $a_{22} = 98.98 a_0$ \cite{Widera2006}. Initially, we set the parameters $a_{12}$, two- and three-body loss terms to zero to obtain insights into the dynamics of Ramsey interferometry.

\begin{figure}[h!]
	\centering
	\includegraphics[width=0.8\linewidth]{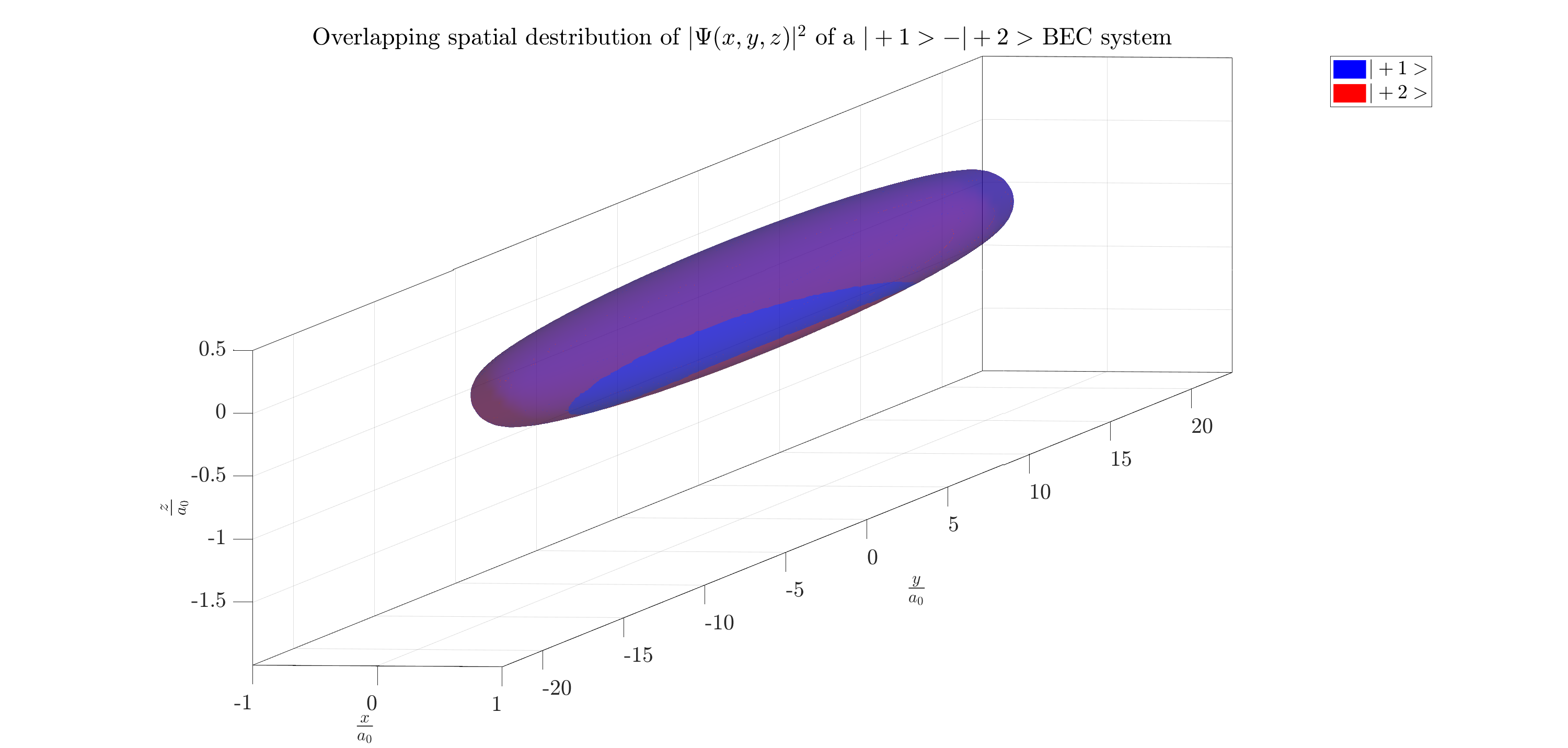}
	\caption[Equal superposition of states of a $\ket{+1}-\ket{+2}$ BEC system]{The spatially overlapping two-component BEC profiles after the first $\frac{\pi}{2}$ Rabi pulse.}
	\label{fig:ASTBEC_50_50}
\end{figure}

For these conditions, first the ground-state for the $\ket{+2}$ BEC was simulated. To create a \num{50}-\num{50} superposition, the two trapping potentials required an energy gap of $\Delta\bar{\Omega}_{12} = \SI{64}{}$ to bring into resonance. This relates to a detuning of $\Delta \approx 2\pi \times \SI{2.4}{\kHz}$. Further, a high Rabi frequency of \SI{10}{\kHz} was selected. This is to minimize the loss of atoms via leaking to the $\ket{+1}$ potential during the Rabi oscillations. Based on this, a \num{50}-\num{50} superposition was created with a resonant Rabi pulse of duration $\bar{t}_R = 0.00189$ corresponding to an actual pulse duration of $t_R = \SI{25}{\us}$ for the chosen trap frequencies. Figure \ref{fig:ASTBEC_50_50} shows the two-component BEC once this stage was reached. Thereafter, several separate scenarios were simulated with the first to verify the Ramsey signal and others for the standard Ramsey sequence.

\begin{figure}[h!]
	\centering
	\includegraphics[width=0.8\linewidth]{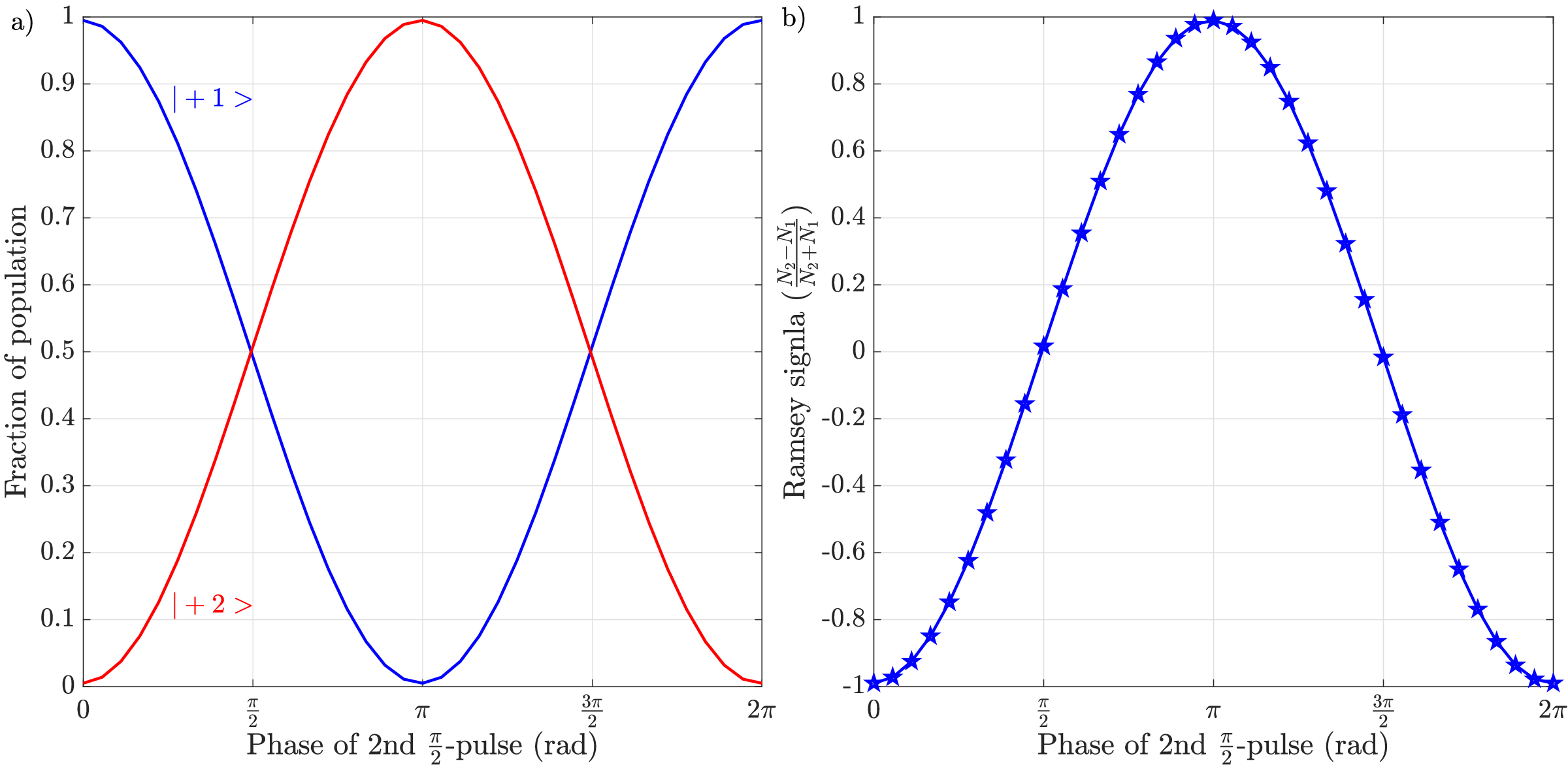}
	\caption[Population variation and Ramsey signal in the $\ket{+1}-\ket{+2}$ BEC system when $a_{12} = 0$ with immediate second $\frac{\pi}{2}$-pulse]{Ramsey interference of the $\ket{+1}-\ket{+2}$ BEC system when $a_{12} = 0$ with immediate second $\frac{\pi}{2}$-pulse. \textbf{a)} Population variation for $\ket{+1}$ (blue) and $\ket{+2}$ (red) BECs and \textbf{b)} Ramsey signal in the phase domain.}
	\label{fig:ASTa0ImmRam}
\end{figure}

In the first scenario, a second recombination pulse with varying phase was applied immediately after the superposition was created. Ramsey fringes (Figure \ref{fig:ASTa0ImmRam} \textbf{b)}) exhibit an interference contrast of \SI{100}{\%} and a full inversion of the populations for the zero phase of the recombination pulse. This is the expected result confirming the capability of the simulation to perform Ramsey interferometry in the phase domain.

\begin{figure}[h!]
	\centering
	\includegraphics[width=0.8\linewidth]{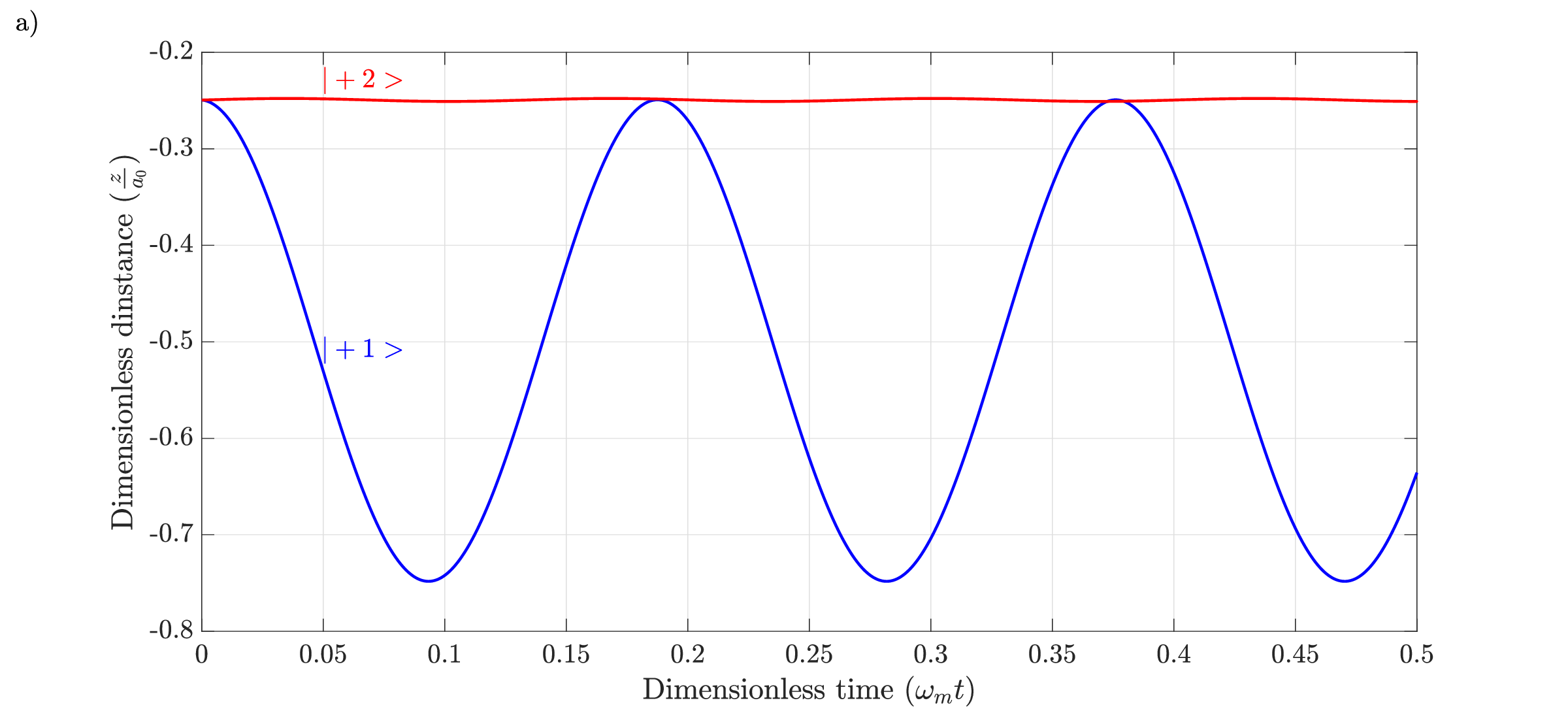}
	\includegraphics[width=0.8\linewidth]{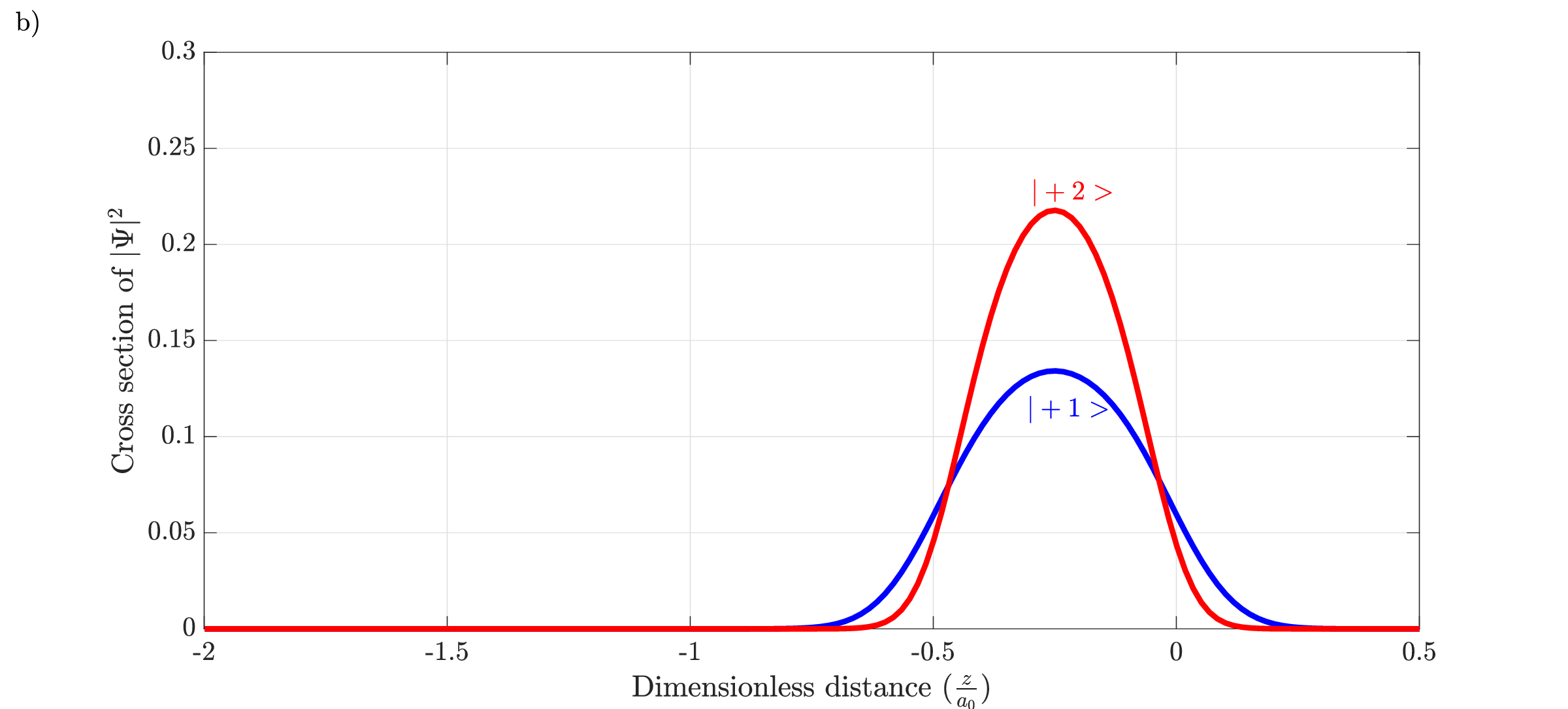}
	\caption[Centre of mass positions of $\ket{+1}-\ket{+2}$ BEC system when $a_{12}=\SI{0}{}$]{a) Centre of mass positions after the first $\frac{\pi}{2}$-pulse of $\ket{+1}$(blue) and $-\ket{+2}$ (red) with $a_{12} = 0$. b) Overlap of condensate wavefunctions $\vert \Psi_1 \rvert^2$ (blue)and $\lvert \Psi_2 \rvert^2$ (red) at the end of the first dipole oscillation in the case of $a_{12} = 0$.}
	\label{fig:ASTa0CoM}
\end{figure}

\pagebreak
The second simulated scenario was to check on dipole oscillations and to observe key physical phenomena that a $\ket{+1}-\ket{+2}$ BEC system may exhibit during the free evolution time. The centre of mass oscillations are shown in Figure \ref{fig:ASTa0CoM} \textbf{a)} and as expected the $\ket{+2}$ BEC remained stationary at its trap bottom while the $\ket{+1}$ BEC oscillated back and forth. The oscillation period was \num{0.1882} in dimensionless units which relates to a dipole oscillation frequency of $\omega_{DP} \approx 2\pi \times \SI{399.3}{\Hz}$. This is in very good agreement with the $2\pi \times \SI{400}{\Hz}$ value of the $\ket{+1}$ radial frequency. Further, the expected idea is for the $\ket{+1}$ BEC to return to the starting spatial point where it overlaps the $\ket{+2}$ BEC. At this point the second $\frac{\pi}{2}$-pulse in the phase domain can be applied to obtain the Ramsey signal. However, as shown in Figure \ref{fig:ASTa0CoM} \textbf{b)}, the overlap region of the two components is reduced.

This is due to the settling of either wavefunction to its new chemical potential during the free evolution time causing dynamical variation of the full-width at half maximum (FWHM). At the end of the first $\frac{\pi}{2}$-pulse, the atom number in $\ket{+2}$ is reduced by one half. This causes its chemical potential to reduce to $\approx \SI{76}{\%}$ of its original value and the theoretical TF radius in the radial direction reduces to $\approx \SI{87}{\%}$ of its original value. The atoms in $\ket{+1}$ experience an expansion of their TF radius due to the relaxed trapping potential. During the free evolution time atoms in $\ket{+1}$ acquire a larger FWHM. Due to these two conditions, the overlap region reduces at the maximum overlap point. 

\begin{figure}[h!]
	\centering
	\includegraphics[width=0.8\linewidth]{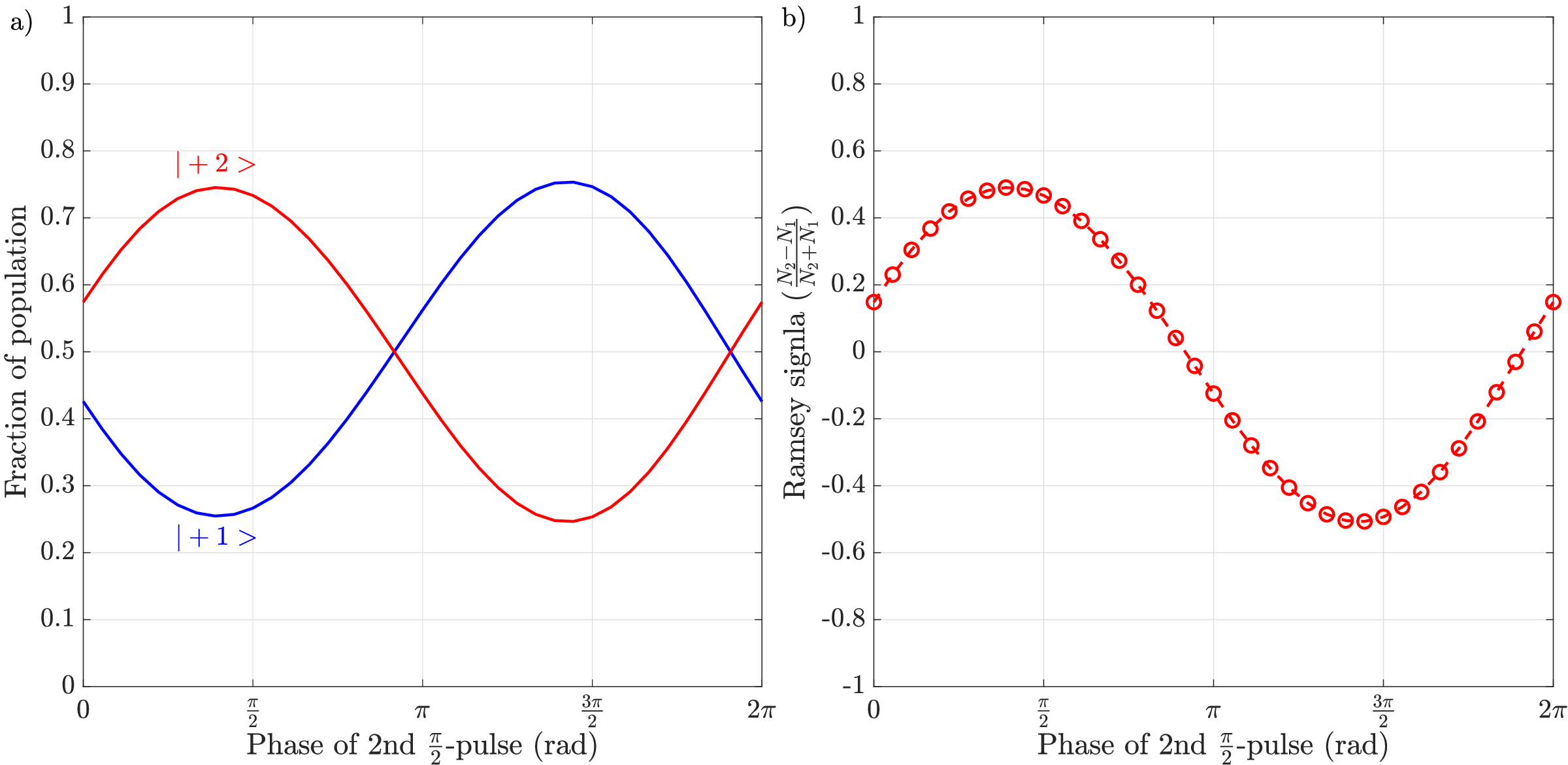}
	\caption[Population variation and Ramsey signal in the $\ket{+1}-\ket{+2}$ BEC system when $a_{12} = 0$ with second $\frac{\pi}{2}$-pulse after one dipole oscillation]{Ramsey interference of the $\ket{+1}-\ket{+2}$ BEC system when $a_{12} = 0$ with second $\frac{\pi}{2}$-pulse applied after one dipole oscillation. \textbf{a)} Population variation for $\ket{+1}$ (blue) and $\ket{+2}$ (red) BECs and \textbf{b)} Ramsey signal in the phase domain.}	
	\label{fig:ASTa0DPRam}
\end{figure}

The reduction in the overlap region as shown in Figure \ref{fig:ASTa0DPRam} reduces the Ramsey contrast as depicted in Equation \ref{eqn:RamseyCon} which prevents the full inversion of populations. When the Ramsey sequence is done the $\ket{+1}-\ket{+2}$ BEC system with $a_{12}=0$ after one dipole oscillation, the resulting Ramsey contrast reduces to $\approx \SI{50}\%$ of its full value as shown in Figure \ref{fig:ASTa0DPRam}. This is an important find as it shows a reduction in the Ramsey contrast even when the inter-state scattering length $a_{12} = 0$, where the contrast reduction mechanism is via the reduced overlap between the two states after one dipole oscillation of $\ket{+1}$.

\pagebreak
The next scenarios consider the impact of the inter-state scattering length on the Ramsey signal. The two considered cases are when $a_{12} = 98 a_0$ and $a_{12} = -50 a_0$ where the focus is the positive and negative nature of $a_{12}$. The initial half of the simulation was repeated for these new configurations in both scenarios and the behaviour of the overlap region is observed. 

\begin{figure}[h!]
	\centering
	\includegraphics[width=0.8\linewidth]{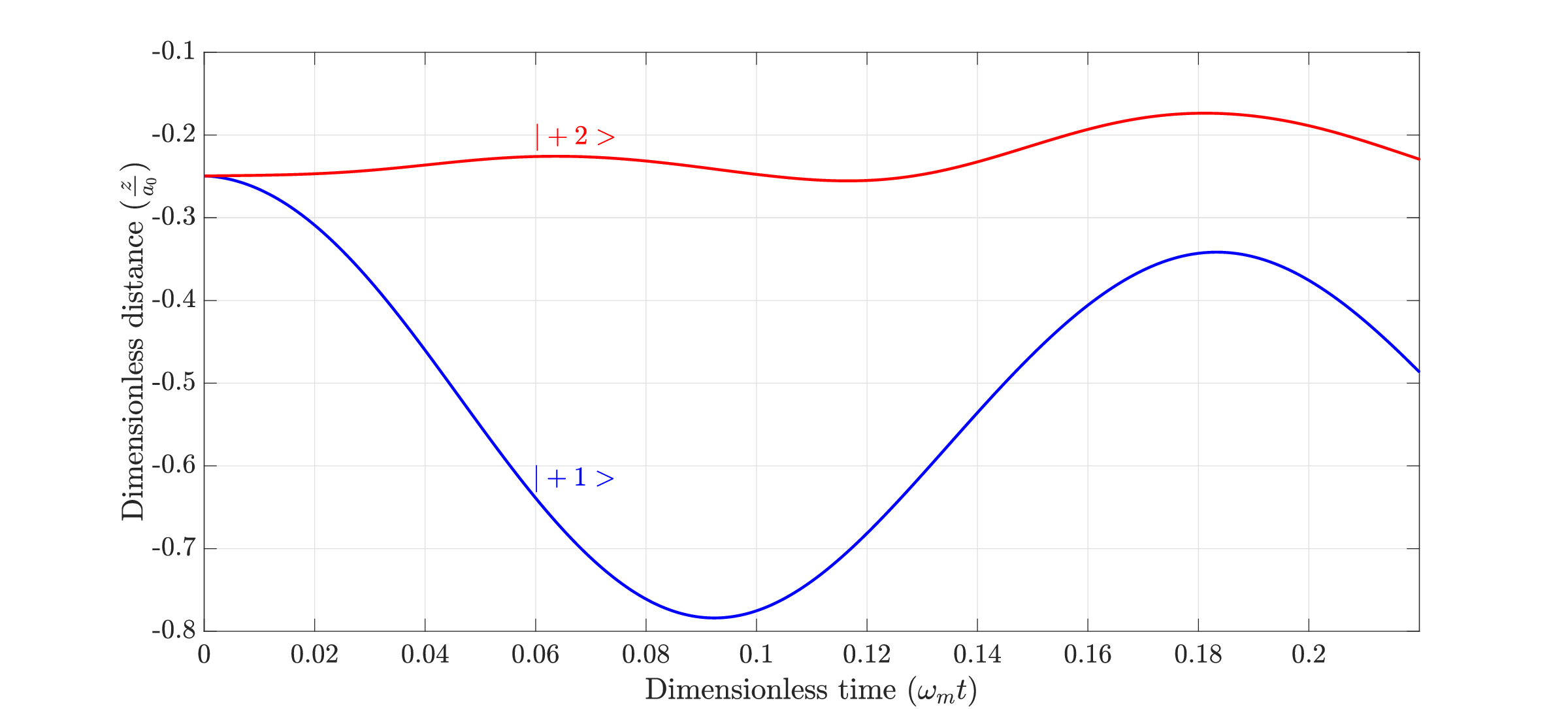}
	\caption[Centre of mass positions of $\ket{+1}-\ket{+2}$ BEC system when $a_{12}=98 a_0$]{Centre of mass positions after the first $\frac{\pi}{2}$-pulse of $\ket{+1}$(blue) and $-\ket{+2}$ (red) with $a_{12} = 98 a_0$.}
	\label{fig:ASTa98CoM}
\end{figure}

\begin{figure}[h!]
	\centering
	\includegraphics[width=0.8\linewidth]{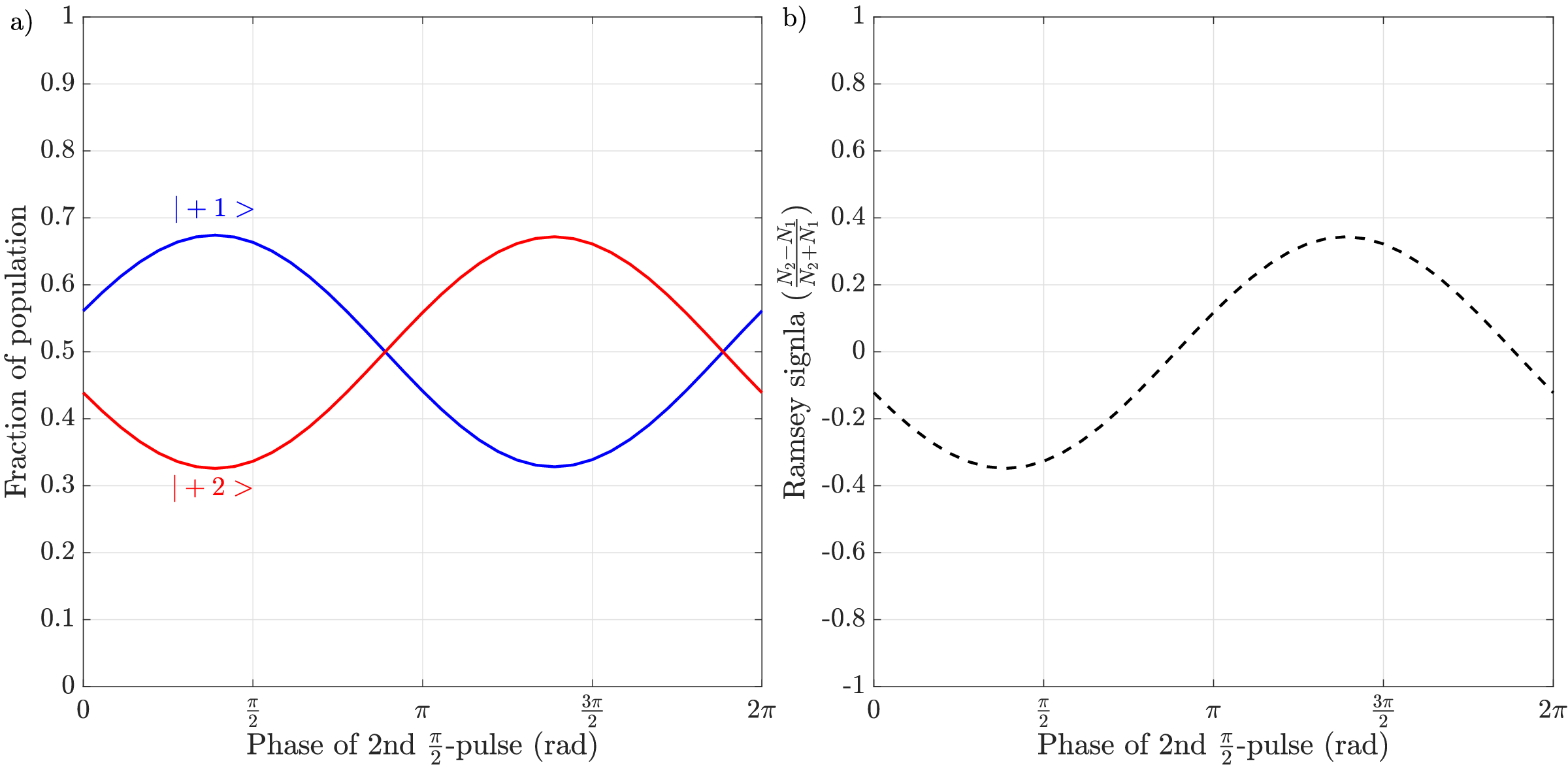}
	\caption[Population variation and Ramsey signal in the $\ket{+1}-\ket{+2}$ BEC system when $a_{12} = 98 a_0$ with second $\frac{\pi}{2}$-pulse after one dipole oscillation]{Ramsey interference of the $\ket{+1}-\ket{+2}$ BEC system when $a_{12} = 98 a_0$ with second $\frac{\pi}{2}$-pulse applied after one dipole oscillation. \textbf{a)} Population variation for $\ket{+1}$ (blue) and $\ket{+2}$ (red) BECs and \textbf{b)} Ramsey signal in the phase domain.}
	\label{fig:ASTa98DPRam}
\end{figure}

The main finding for $a_{12} = \SI{98} a_0$ is reflected in the centre of mass motions of the two states during the dipole oscillations. As shown in Figure \ref{fig:ASTa98CoM}, during the first separation the $\ket{+1}$ BEC exerts a force that pushes the $\ket{+2}$ BEC away from its rest point. This force originates from the non zero inter-state scattering length which makes the two states susceptible to each other. Further, the two BECs repel each other away when $\ket{+1}$ BEC returns back from one dipole oscillation which is clearly shown at \num{0.18} dimensionless time units in Figure \ref{fig:ASTa98CoM}. This leads to a further reduction in the overlap region between the two states at the recombination pulse. Due to this the population does not undergo full inversion at the second $\frac{\pi}{2}$-pulse (Figure \ref{fig:ASTa98DPRam} \textbf{a)}) and the Ramsey signal decreases (Figure \ref{fig:ASTa98DPRam} \textbf{b)}) to about $\SI{35}\%$ of its full value.

\begin{figure}[h!]
	\centering
	\includegraphics[width=0.8\linewidth]{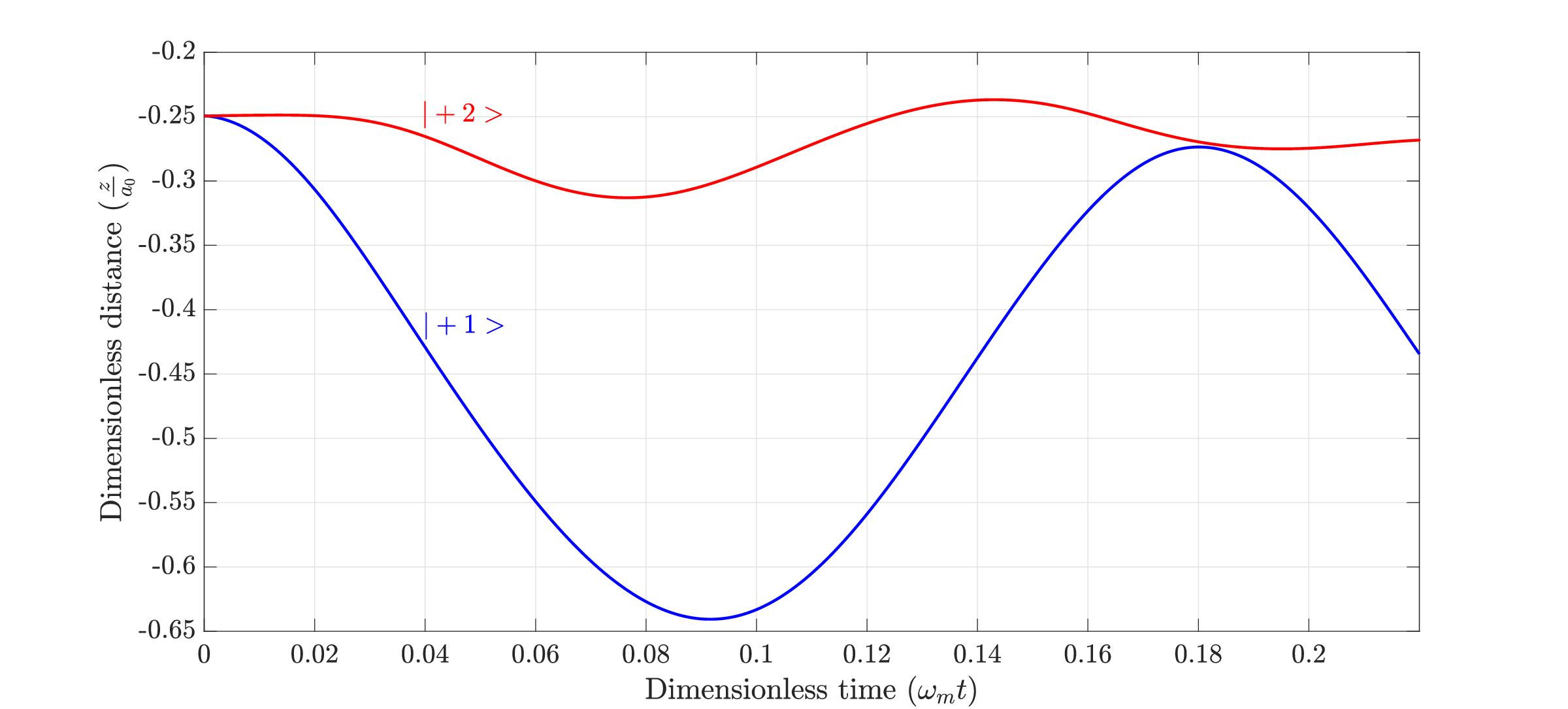}
	\caption[Centre of mass positions of $\ket{+1}-\ket{+2}$ BEC system when $a_{12}=-50 a_0$]{Centre of mass positions after the first $\frac{\pi}{2}$-pulse of $\ket{+1}$(blue) and $-\ket{+2}$ (red) with $a_{12} = -50 a_0$.}
	\label{fig:ASTaN50CoM}
\end{figure}

\begin{figure}[h!]
	\centering
	\includegraphics[width=0.8\linewidth]{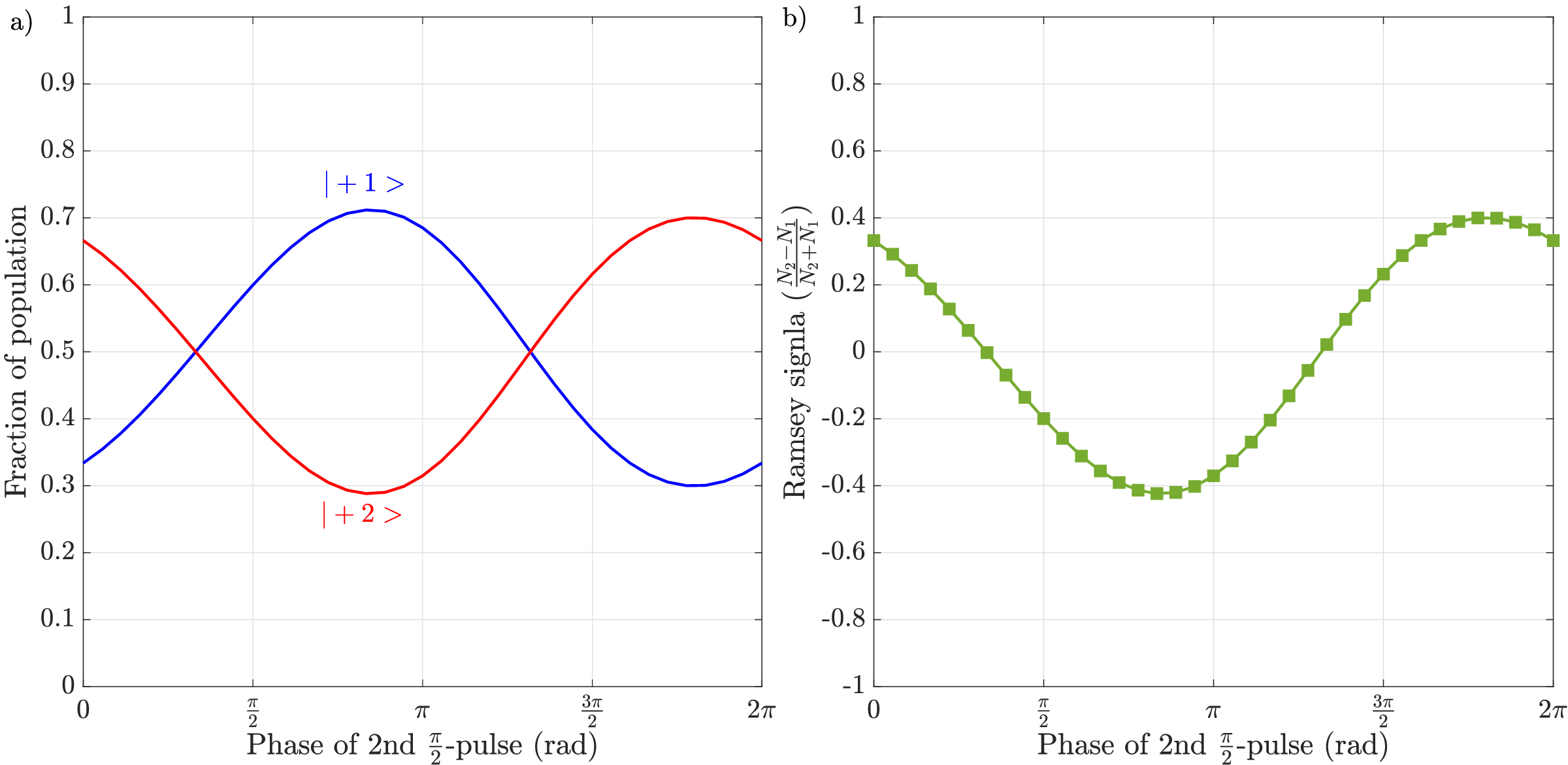}
	\caption[Population variation and Ramsey signal in the $\ket{+1}-\ket{+2}$ BEC system when $a_{12} = -50 a_0$ with second $\frac{\pi}{2}$-pulse after one dipole oscillation]{Ramsey interference of the $\ket{+1}-\ket{+2}$ BEC system when $a_{12} = -50 a_0$ with second $\frac{\pi}{2}$-pulse applied after one dipole oscillation. \textbf{a)} Population variation for $\ket{+1}$ (blue) and $\ket{+2}$ (red) BECs and \textbf{b)} Ramsey signal in the phase domain.}
	\label{fig:ASTaN50DPRam}
\end{figure}

\pagebreak
When $a_{12} = -50a_0$, the two-component system exhibits attractive interactions which is clearly reflected in Figure \ref{fig:ASTaN50CoM}. Here, the centre of mass motion of $\ket{+1}$ pulls on $\ket{+2}$ during the initial stage of the dipole oscillation and the effect is reversed when the overlap occurs at about \num{0.18} dimensionless time units. This has an impact on the overlap of the two components at the end of one dipole oscillation which is reflected in the Ramsey signal (Figure \ref{fig:ASTaN50DPRam} \textbf{b)}) which is reduced to $\approx \SI{45}\%$ of its full value. This is better than the case of $a_{12}=98 a_0$, but not as good as the best possible contrast of \SI{50}{\%} for the zero inter-state scattering length.

\begin{figure}[h!]
	\centering
	\includegraphics[width=0.8\linewidth]{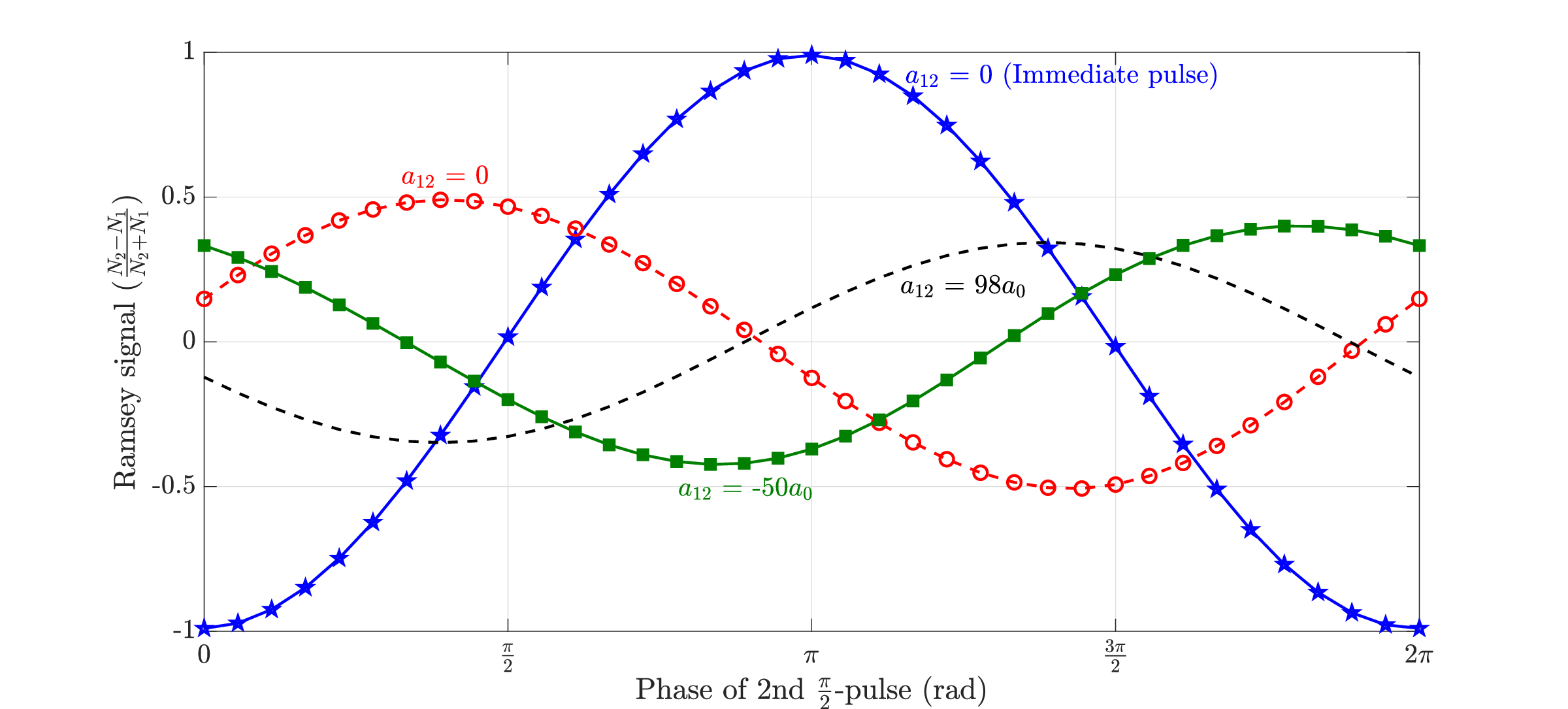}
	\caption[Different Ramsey fringes in the $\ket{+1}-\ket{+2}$ BEC system]{Different Ramsey fringes of the $\ket{+1}-\ket{+2}$ BEC system, where the blue solid line with stars is for $a_{12}=0$ with immediate recombination pulse, red dashed line with open circles is for $a_{12}=0$ with recombination pulse at one dipole oscillation, black dashed line is for $a_{12}=98a_0$ and green solid line with solid squares is for $a_{12}=-50a_0$.}
	\label{fig:ASTRamComp}
\end{figure}

\pagebreak
Finally, Figure \ref{fig:ASTRamComp} compares the Ramsey signals between the various scenarios considered above. The blue solid line with stars is when $a_{12}=0$, where the recombination pulse is applied immediately after the first pulse in the Ramsey sequence when both states do not move. This has a phase offset of \SI{4.7}{\radian} with Ramsey signal at \SI{100}{\%}. The red dashed line with open circles is when $a_{12}=0$, where the recombination pulse is applied after one dipole oscillation of $\ket{+1}$ in the Ramsey sequence. This has a phase offset of \SI{0.3}{\radian} with Ramsey signal at about \SI{50}{\%}. The particular feature is that $\ket{+2}$ does not move during this sequence and only $\ket{+1}$ moves (Figure \ref{fig:ASTa0CoM} \textbf{a)}) which contributes to the phase offset compared to the in-trap scenario. The next case is the black dashed line when $a_{12}=98 a_0$, where the recombination pulse is applied after one dipole oscillation of $\ket{+1}$ in the Ramsey sequence. This has a phase offset of \SI{3.5}{\radian} with the Ramsey signal at about \SI{35}{\%}. Here, the $\ket{+2}$ also moves during this sequence (Figure \ref{fig:ASTa98CoM}) which contributes to the phase offset of about \SI{3.2}{\radian} compared to $a_{12}=0$ with one dipole oscillation (red dashed line with open circles). Finally, the green solid line with squares when $a_{12}= -50 a_0$, where the recombination pulse is applied after one dipole oscillation of $\ket{+1}$ in the Ramsey sequence. This has a phase offset of \SI{2.0}{\radian} with Ramsey signal at about \SI{45}{\%}. Here, the $\ket{+2}$ moves the most out of all the cases during the Ramsey sequence (Figure \ref{fig:ASTaN50CoM}) which contributes to the phase offset of about \SI{1.7}{\radian} compared to $a_{12}=0$ with one dipole oscillation (red dashed line with open circles). 

\section*{Discussion}
The above shows intriguing results for the $\ket{+1}-\ket{+2}$ BEC system and several contrast reduction mechanisms for the Ramsey signal. It should be noted that, though the amplitude of the Ramsey signal is small, the integrity of the signal over several dipole oscillations may be explored depending on experimental findings. All in all, the above displays the capabilities of the developed tool which is a good platform to explore and interpret experimental data.

\bibliography{References}







\end{document}